\documentclass[acmsmall,screen]{acmart}

\AtBeginDocument{%
  \providecommand\BibTeX{{%
    \normalfont B\kern-0.5em{\scshape i\kern-0.25em b}\kern-0.8em\TeX}}}

\setcopyright{acmlicensed}
\acmJournal{TOIS}
\acmYear{2023}
\acmDOI{10.1145/3637869}

\usepackage{bbm}
\usepackage{subfigure}
\usepackage{caption}
\usepackage{xspace}
\usepackage{ulem}
\usepackage{makecell}
\usepackage[T1]{fontenc}    

\usepackage{amsmath,amsfonts,bm, multirow}

\def\eqref#1{equation~\ref{#1}}

\def\1{\bm{1}}

\DeclareMathAlphabet{\mathsfit}{\encodingdefault}{\sfdefault}{m}{sl}
\SetMathAlphabet{\mathsfit}{bold}{\encodingdefault}{\sfdefault}{bx}{n}

\DeclareMathOperator*{\argmax}{arg\,max}

\newcommand{\fig}[1]{Fig.~\ref{#1}}
\newcommand{\eq}[1]{Eq.~(\ref{#1})}
\newcommand{\tb}[1]{Tab.~\ref{#1}}
\newcommand{\se}[1]{Section~\ref{#1}}

\newcommand{\defi}[1]{Definition~\ref{#1}}

\newcommand{\bbE}{\ensuremath{\mathbb{E}}} 
\newcommand{\caM}{\ensuremath{\mathcal{M}}} 
\newcommand{\caD}{\ensuremath{\mathcal{D}}} 
\newcommand{\caL}{\ensuremath{\mathcal{L}}} 
\newcommand{\caF}{\ensuremath{\mathcal{F}}}

\newcommand{\caR}{\ensuremath{\mathcal{R}}}

\newcommand{\our}{\textsc{Mirror}\xspace}

\newtheorem{assumption}{Assumption}

\begin{document}

\title{Understanding or Manipulation: Rethinking Online Performance Gains of Modern Recommender Systems}

\author{Zhengbang Zhu}
\affiliation{%
  \institution{Shanghai Jiao Tong University}
  \country{China}}
\email{zhengbangzhu@sjtu.edu.cn}

\author{Rongjun Qin}
\affiliation{%
  \institution{National Key Laboratory for Novel Software Technology, Nanjing University}
  \country{China}}
\affiliation{%
  \institution{Polixir Technologies}
  \country{China}}
\email{qinrj@polixir.ai}

\author{Junjie Huang}
\affiliation{%
  \institution{Shanghai Jiao Tong University}
  \country{China}}
\email{legend0018@sjtu.edu.cn}

\author{Xinyi Dai}
\affiliation{%
  \institution{Shanghai Jiao Tong University}
  \country{China}}
\email{daixinyi@sjtu.edu.cn}

\author{Yang Yu$^\dagger$}
\affiliation{%
  \institution{National Key Laboratory for Novel Software Technology, Nanjing University}
  \country{China}}
\affiliation{%
  \institution{Polixir Technologies}
  \country{China}}
\email{yuy@polixir.ai}

\author{Yong Yu}
\affiliation{%
  \institution{Shanghai Jiao Tong University}
  \country{China}}
\email{yyu@apex.sjtu.edu.cn}

\author{Weinan Zhang$^\dagger$}
\affiliation{%
  \institution{Shanghai Jiao Tong University}
  \country{China}}
\email{wnzhang@sjtu.edu.cn}

\thanks{$^\dagger$Corresponding authors}

\renewcommand{\shortauthors}{Z. Zhu et al.}

\begin{abstract}

Recommender systems are expected to be assistants that help human users find relevant information automatically without explicit queries. As recommender systems evolve, increasingly sophisticated learning techniques are applied and have achieved better performance in terms of user engagement metrics such as clicks and browsing time. The increase in the measured performance, however, can have two possible attributions: a better understanding of user preferences, and a more proactive ability to utilize human bounded rationality to seduce user over-consumption. A natural following question is whether current recommendation algorithms are manipulating user preferences. If so, can we measure the manipulation level? In this paper, we present a general framework for benchmarking the degree of manipulations of recommendation algorithms, in both slate recommendation and sequential recommendation scenarios. The framework consists of four stages, initial preference calculation, training data collection, algorithm training and interaction, and metrics calculation that involves two proposed metrics, Manipulation Score and Preference Shift. We benchmark some representative recommendation algorithms in both synthetic and real-world datasets under the proposed framework. We have observed that a high online click-through rate does not necessarily mean a better understanding of user initial preference, but ends in prompting users to choose more documents they initially did not favor. Moreover, we find that the training data have notable impacts on the manipulation degrees, and algorithms with more powerful modeling abilities are more sensitive to such impacts. The experiments also verified the usefulness of the proposed metrics for measuring the degree of manipulations. We advocate that future recommendation algorithm studies should be treated as an optimization problem with constrained user preference manipulations.
\end{abstract}



\maketitle

\section{Introduction}

With the popularity of the Internet and the growth of User Generated Content (UGC)~\cite{krumm2008user} during recent years, we are being overwhelmed by a massive volume of information. To save users from information overload, recommender systems have been widely applied in today's short video~\cite{liu2019user}, news~\cite{wang2018dkn} and e-commerce~\cite{chen2019behavior} platforms. 
Different from traditional information retrieval techniques, recommender systems are considered to be more advanced by using user information to personalize the recommendations. This is accomplished through the prediction of the relevance between users and documents.
There are various recommendation algorithms proposed for relevance modeling, and their development is evidenced by the progressive improvement in various offline and online metrics. Offline metrics refer to those that can be evaluated with a static dataset and without interactions with real users. For example, Normalized Discounted Cumulative Gain (NDCG) is commonly used in recommendation literature to measure how well the model ranks more relevant documents at the top of the list. On the other hand, online metrics have to be computed by deploying the trained model online and gathering feedback from real users. As a general case of online metrics, Click-Through Rate (CTR) shows how often people click on the displayed documents by a recommendation algorithm. 

While improved online metrics bring more traffic and revenue to the platform, it is questionable whether existing online metrics are aligned with the aspiration of recommender systems, which is to help people find relevant or favored content. As revealed by research in behavioral economics, humans make decisions with \textit{bounded rationality}~\cite{selten1990bounded}. Bounded rationality is contrary to expected-utility models~\cite{schoemaker1982expected} where decisions made are always optimal under some expected utility function. Instead, humans are trying to make optimal decisions, but are bounded by various cognitive limitations, e.g., limited memory and decision time. Under the bounded rationality framework, a variety of phenomena have been observed and investigated, such as the decoy effect, confirmation bias, anchoring effect, etc~\cite{slaughter1999decoy, nickerson1998confirmation, furnham2011literature}. Most of them are possible in online recommendations, which are already studied by prior works~\cite{teppan2011decoy, lex2018mitigating, adomavicius2018effects}. Once a recommender system is deployed online, it will actively interact with human users of bounded rationality. Therefore, we cannot distinguish whether the active participation of users is due to the system's accurate grasp of their preferences or to an over-exploitation of their psychological weaknesses. To our knowledge, there is no existing work that studies whether the performance gain of current recommendation algorithms in online metrics comes from the utilization of bounded rationality.

Since intrusive evaluations directly on the production platform can have a negative impact on the company's benefits, building simulated interactive environments is used to assess the properties of algorithms when interacting with users without actually going online. Agent-based user simulations are adopted by early works in building dialogue systems~\cite{schatzmann2006survey}, where user behaviors are simulated by either predefined rules~\cite{levin2000stochastic, pietquin2006probabilistic} or statistical models trained on a small amount of data~\cite{lemon2006evaluating}. In recent years, a series of simulation frameworks or specific environments have been proposed in the field of recommender systems. Recsim~\cite{ie2019recsim} is a general framework for simulating the interactive process of online recommendations, where the whole recommender system is decomposed into separate configurable components. Shi et.al.~\cite{shi2019virtual} propose Virtual-Taobao as an interactive e-commence environment, in which the user behaviors are given by a neural model adversarially trained on real logged data. However, the evaluation criteria adopted in existing works are aligned with traditional online metrics, and there is no simulation framework to specifically test the extent to which recommendation algorithms manipulate user preferences.

\begin{figure}[t]
\centering
\includegraphics[width=0.9\textwidth]{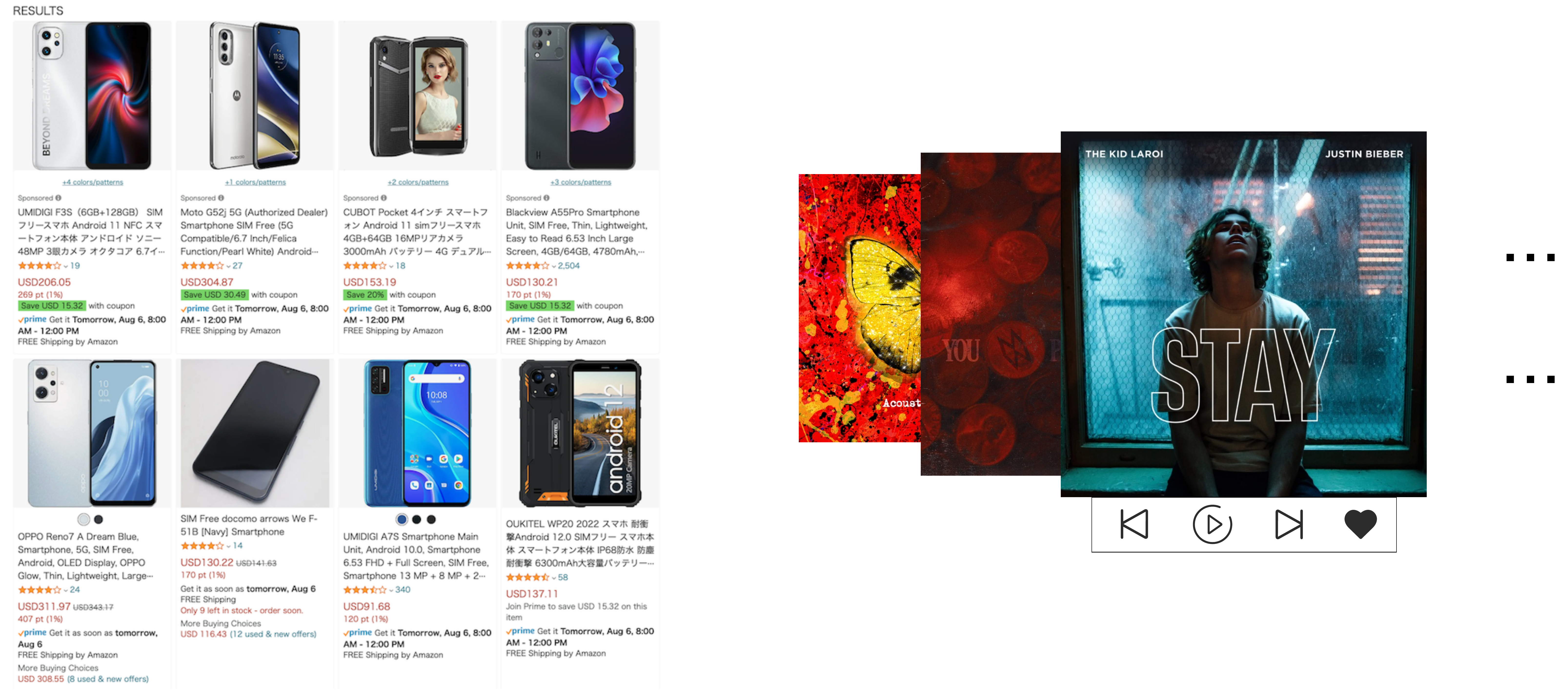}
\caption{Examples of two typical scenarios in recommender systems. On the left is the search result in an e-commerce platform, referred to as slate recommendation. On the right is the illustration of a recommendation queue in music platforms, referred to as sequential recommendation.}
\label{fig:recsys-example}
\end{figure}

To relieve these concerns, in this paper, we provide a general simulation framework \our to evaluate the manipulations of user preferences from recommendation algorithms. The evaluation consists of four stages. 
In the first stage, each simulated user is presented with all the documents one by one to get a score for each document. Since the scoring happens before the interactions with recommender systems, and each document itself makes up a slate, such scores are not influenced by historically viewed documents or by other documents within the same recommended slate. Therefore, the obtained scores can serve as the initial preferences of users. 
In the second stage, we generate training datasets by using pre-defined initial recommendation strategies to interact with users. 
To acquire a diverse set of training datasets to study the impact of how training datasets influence trained models, we mix the datasets from multiple initial strategies with different ratios. 
After that, in the third stage, the recommendation model is trained on those datasets and evaluated by interacting with simulated users. 
At the final stage, the interaction data, as well as users' initial preferences, are jointly used to compute evaluation metrics. By incorporating the initial preferences, we define the notion of favorite documents and propose several manipulation-aware metrics. Those metrics can be used in both slate and sequential recommendations, which are two typical recommendation scenarios as illustrated in \fig{fig:recsys-example}. Intuitively, if the users click on more documents but less proportion of favorite documents, we say the recommendation algorithms are manipulating the users' preferences. Also, the manipulations can be quantified by the long-term shift in users' unbiased preferences. 

Under the proposed framework, we conduct extensive studies based on four different simulation experiments, two of which are rule-based and the other two are data-driven. Those environments cover both slate and sequential recommendations. In slate recommendations, we find that compared to simple point-wise models, the reranking methods with the ability to model item relationships within one slate tend to generate recommendations that have higher clicks but a lower proportion of clicks on favorite documents. As the training data contains more slates with manipulations, all the algorithms recommend slates with increasing clicks and decreasing proportion of favorite documents clicked. Moreover, our case study reveals that the reranking method makes use of comparison bias, where people tend to choose those documents if they have already seen similar but worse alternatives within the same list. In sequential recommendations, the experimental results show that sequential recommendation algorithms that take users' recent interaction history into account can lead to more significant manipulation of users' preferences.

To summarize, the main contributions of our work are:
\begin{itemize}
    \item We put forward the potential issue of manipulations on user preference in current recommendation systems, and propose a general and configurable framework named \our that can quantify the degree of such manipulations of recommendation algorithms. The core procedure inside the benchmark is to make simulated users view and score the entire documents one by one, through which we can access the initial preferences of each user.
    \item By utilizing the proposed framework, we instantiate four benchmark scenarios, ranging from slate recommendations and sequential recommendations, and including both synthetic and data-driven user behavior models. 
    \item We benchmark several recommendation algorithms under these scenarios, and draw up a number of key findings on manipulations from recommendation systems.
    \begin{itemize}
        \item In both slate and sequential recommendations, better performances on traditional online metrics are accompanied by more manipulations on user preference. 
        \item In both slate and sequential recommendations, the degree of manipulation from recommendation algorithms is overall positively correlated with the manipulation degree of the recommendation algorithm that collects the training data.
        \item In slate recommendations, reranking methods are more actively using the mutual influence of documents within the slate to improve the overall clicks compared to point-wise ranking methods. And such influence often leads users to click on documents they do not originally favor.
        \item In sequential recommendations, recommendation models that take the user's recent interaction behavior into account can make users choose less proportion of originally favored documents and induce greater changes to user preferences.
    \end{itemize}
\end{itemize}

The remaining part of this paper is organized as follows. \se{sec:overview} provides an overview of current recommender systems and the emerging concerns on negative impacts.
In \se{sec:assumption}, we make two necessary assumptions to delimit the scope of manipulations studied in this paper.
In \se{sec:framework}, we introduce our benchmark framework \our, including its components, benchmark procedure and evaluation metrics. In \se{sec:exp}, we conduct four benchmark experiments and analyze the results. Related works are summarized in \se{sec:related}. We finally conclude this paper in \se{sec:conclusion}.

\section{An Overview of Current Recommender Systems}
\label{sec:overview}

We are living in an era of information explosion. With the popularity of the internet, we have access to far more information than ever before, making it difficult to find what we need among the vast amount of content. 
To save people from information overload and find the most relevant content for each of us, modern recommender systems make use of rich user profiles to model user preferences. These systems utilize techniques like collaborative filtering to guess what users might favor, using not only their own but also other users' behaviors.
The more you use the recommender system, and the more other people are using the system, the more you will be recommended with relevant results \cite{sarwar2001item,koren2008factorization}.

If we look at the other side, the situation is also thriving. For companies who are deploying large-scale recommender systems on their products or platforms, personalized recommendation greatly improves the retention rate of users, as well as the clicks or purchases, depending on the service they provide. The method from a pioneer study~\cite{das2007google} from Google achieves an average increase of 38\% in click rates on its news platforms with the personalized recommendation, compared to general methods. Even more gratifyingly, the e-commerce giant e-Bay reported a 500\% spike in Gross Merchandise Volume in online A/B Testing~\cite{chen2011recommending} with their personalized recommendation. In industry, recommender systems are continuing to iterate, taking advantage of the increasing amount of data and computing power available to deliver significant revenue for businesses.

\begin{figure}[h]
\centering
\includegraphics[width=0.9\textwidth]{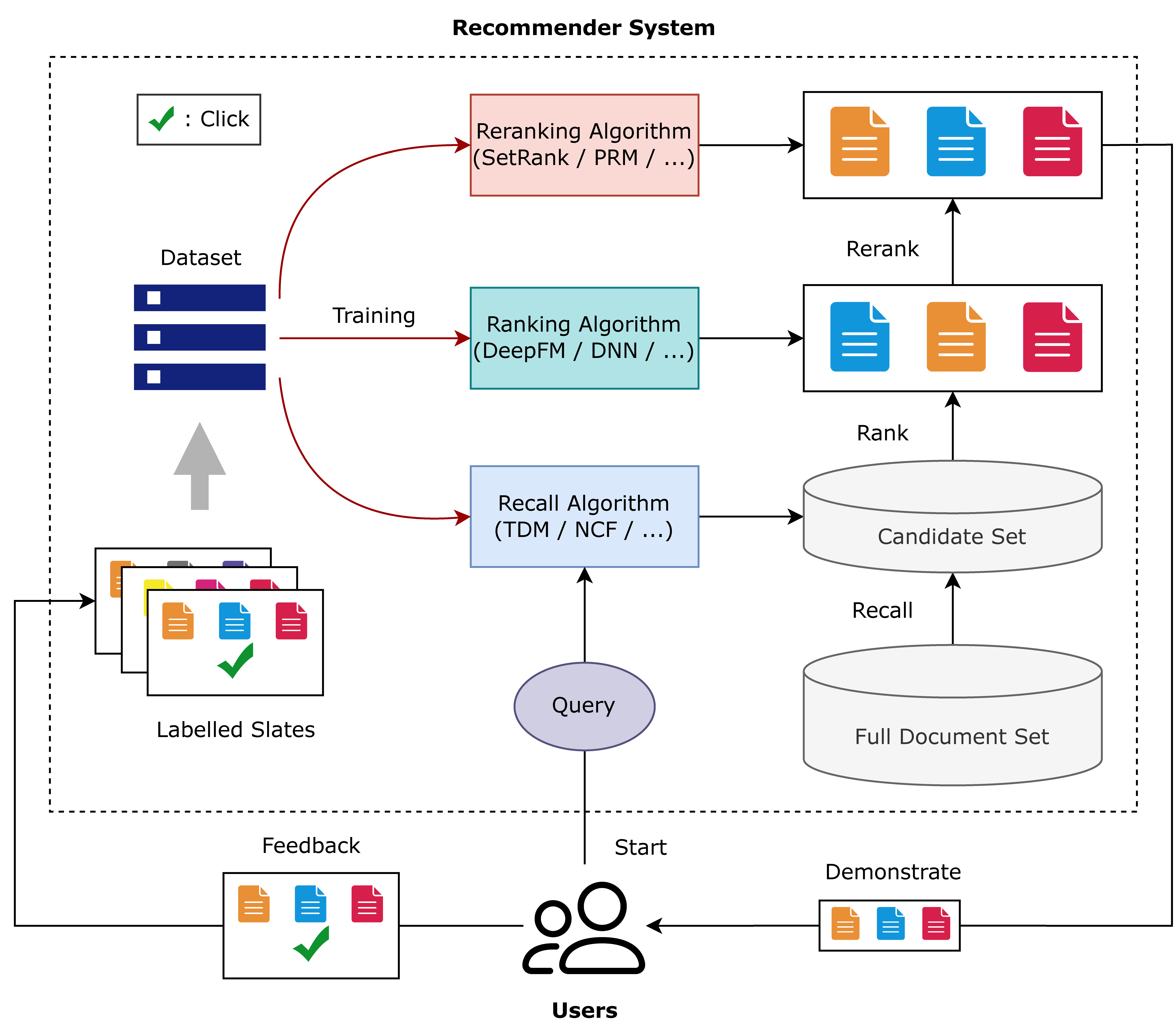}
\caption{The overview of the workflow of a typical recommender system. Here we show user feedback as clicks, but in practice, user feedback can also be purchases, ratings, or just browsing, etc.}
\label{fig:recsys-overview}
\end{figure}

Let us briefly review the working mechanism of a standard recommender system. As illustrated in \fig{fig:recsys-overview}, we consider the multi-stage recommendation setting with user clicks as feedback. The users can interact with the recommender system $\caR$ for one or multiple rounds. 
Each round starts with a user $u$ proposing a query $q$ to $\caR$. 
During round $r$, the user is recommended with a sorted list of $T$ documents $D^r=\{d^r_1, d^r_2, \dots, d^r_T\}$, and the user can choose to click on any number of documents, forming a click sequence $C^r=\{c^r_1, c^r_2, \dots, c^r_T\}$. The process continues until the user exits.

The recommended list $D^r$ is generated by the recommender system from a three-stage procedure. A small candidate set $\caD_q$ is first retrieved based on $q$ by a \textit{recall algorithm} from the full document set. Then the documents in the candidate set are scored and sorted by a \textit{ranking algorithm}. Top-$T$ ranked items form the initial ranking list and are subsequently processed by a \textit{reranking algorithm} to get a refined order. The reordered list is used as the recommended list $D^r$ and displayed to the user. Feedback from users is logged and stored as the dataset, which is used for improving the algorithms of each stage. The ranking and reranking algorithms are the core of the recommender system and have experienced rapid growth in recent years. Since deep learning-based methods play a dominant role in the industry \cite{cheng2016wide,zhou2018deep,guo2017deepfm}, here in this paper, we only discuss ranking/reranking algorithms that are based on neural networks. 

In the following parts, we consider two different scenarios of recommendation. In the first slate scenario, we only use the user-item information within this round as the input of the recommender. Ranking and reranking algorithms are discussed in this part. In the second sequential scenario, we use cross-round information as model input. Users' historical interaction data from previous rounds are used as model input to capture the users' preference drift. In this part, we include traditional sequential recommendation solutions and recent offline reinforcement learning-based methods. 


\subsection{Slate Recommendation}

Learning to rank (LTR) methods have been widely applied in slate recommendation ~\cite{zhao2019recommending}, where the most relevant items are sorted and displayed as ranked lists. 
Apart from simple point-wise approaches where ranking is modeled as the accurate prediction of the ground truth label for a single document, pairwise or listwise approaches have been explored.
Pairwise approaches consider the relative order between a pair of items. 
As a representative of pairwise methods, RankNet~\cite{burges2005learning} formulates the ranking problem as pairwise classification, where the loss function is constructed by the relative relationship between scores of item pairs.
Listwise approaches consider the entire group of documents in the candidate set and their order in the list.
LambdaRank and ListMLE are two important list-wise approaches. 
LambdaNet~\cite{burges2010ranknet} uses a weighted pairwise loss function to optimize the bound of ranking metrics such as NDCG ~\cite{wang2018lambdaloss}. 
ListMLE~\cite{xia2008listwise} constructs the maximum likelihood loss on items in the entire list with the Plackett-Luce model assumption, which tackles the ranking list generation task in a more direct way.

In real-world ranking systems, neural structures are often used to extract rich feature interactions of each user-item pair. \citet{he2017neural} utilizes a multi-layer perceptron (MLP) structure to model the feature interactions of each user-item pair, which is extended in \cite{he2018outer} by using convolutional neural networks (CNN) to further extract potential interactions across embedding dimensions. Wide \& Deep~\cite{cheng2016wide} is built upon a hybrid network structure, where a linear model delivers a wide set of cross-product features along with a deep feed-forward network modeling complex and implicit interactions. DeepFM~\cite{guo2017deepfm} combines the factorization machine and deep models to handle low-order and high-order separately, and no need for any feature engineering compared to Wide \& Deep.



In Multi-stage Recommender Systems (MRS), besides former deep models that encode each user-item pair separately, neural reranking models are applied after the normal ranking stage to further encode interactions between items.
After candidate items are scored and ranked by ranking models, the top-ranked ones are fed into a sequence model for reranking rather than directly displayed to the user. The reranking model takes the mutual influence of items in the same list into consideration, which can further improve the quality of recommended lists. DLCM~\cite{ai2018learning} employs a recurrent neural network to learn a local context embedding of all top retrieved items from the ranking process. Later, PRM~\cite{pei2019personalized} and SetRank~\cite{pang2020setrank} adopt the self-attention structure for more flexibility and efficiency, where PRM additionally utilizes a pre-trained personalized matrix to calculate the embedding. 

\subsection{Sequential Recommendation}

Sequential recommendation utilizes users' historical interaction data to capture the dynamic drifts in user preferences. Sequential recommendation views users' preferences as an ever-changing process, which can be influenced by previously recommended items or other external factors, and expressed as users' recent behaviors. Earlier works based on Markov chains~\cite{he2016vista, he2016fusing, rendle2010factorizing} model the users' behavior as a Markov chain and predict users' next behavior based on their previous actions. Deep learning-based methods have been popular in sequential recommendation. There are RNN-based methods~\cite{beutel2018latent, hidasi2018recurrent, hidasi2015session,jing2017neural}, CNN-based methods~\cite{kang2018self,tang2018personalized} and transformer-based methods~\cite{kang2018self}.


Offline reinforcement learning (RL) methods are also adopted to facilitate sequential recommendations. Reinforcement learning originated from the optimal control domain, which actively learns control policies from reward signals collected through continual interactions with the environment, e.g., video games~\cite{mnih2013playing} or physical simulators~\cite{schulman2017proximal}. The cost of interacting with the environment is one of the major obstacles to applying RL methods to real-world scenarios. Such cost is especially notable in the recommendation, where the randomness in the recommended items might harm the user experience and reduce the platform revenue. Offline RL methods are RL algorithms that only learn from static datasets without interaction with the user, which enjoy the advantage of RL-based methods but avoid the exploration cost. \citet{xiao2021general} propose a general framework for applying offline RL in sequential recommendations and using a stochastic actor-critic structure with several regularization techniques to stabilize training. In a study by Garg et al.~\cite{garg2020batch}, a variant of Batch-Constrained Q-Learning~\cite{fujimoto2019off} with distributional objective is applied and outperforms the supervised learning methods to a large extent.

\subsection{Rising Concerns Behind the Prosperity}
\label{sec:overview-concerns}
We have witnessed that the evolution of existing recommendation algorithms is mainly along the lines of using increasingly complex neural networks to model, using increasingly rich interaction data, and even employing more active optimization goals (as those used in offline RL methods). On the one hand, recommender systems are developed and deployed by companies and platforms as a tool to improve revenue, and users do not have access to technical details.
On the other hand, due to the black-box nature of neural networks, especially those with numerous parameters, even the developers themselves can not fully understand how the models work~\cite{zhang2021survey, fan2020interpretability}.
Such a lack of transparency has led to rising concerns about the seeming success of recommender systems. There are several recent studies on the ethical issues with the commercial recommendation. A comprehensive review paper~\cite{milano2020recommender} discusses ethical challenges in deploying recommender systems, such as fairness, privacy, and potential negative social impacts. Paraschakis~\cite{paraschakis2017towards} has proposed a framework to guide how to promote fairness, privacy protection, and algorithmic opacity throughout the workflow of the recommendation system. However, how to completely eliminate the aforementioned ethical problems remains an open problem.

We focus on another ethical question that is seldom mentioned by previous works: \textit{are recommender systems manipulating users' preferences for higher revenue?} Typical recommender systems are trained with offline data and evaluated using both offline and online metrics. Offline metrics measure how well the algorithm fits the distribution of the data set, while the situation with online metrics is more complicated. Human decision-making has bounded rationality and is easily influenced by the displayed items. The behavioral economics community has studied some typical phenomena, e.g., the decoy effect, the confirmation bias, and the anchoring effect~\cite{josiam1995consumer, charness2017confirmation, bergman2010anchoring, camerer1999behavioral}. 
Recommendation systems are likely to exploit these psychological weaknesses and biases to achieve higher performance in online metrics, which we call \textit{manipulation}. While such manipulation can lead to an increase in user clicks, the items clicked on are not necessarily those the user preferred in the first place.

The manipulations of user preferences can arise either intentionally or unintentionally. 
Since most recommender systems operate as a ``black box'', it is hard to distinguish whether manipulations originate from intentional design or more sophisticated modeling of user behavior and biases in the data, etc.
We are not trying to explore the causes or motives behind the manipulative behavior of recommender systems, but rather to quantitatively measure whether such manipulation exists. 
From a consequentialist perspective, recommender systems that manipulate user preferences cause equal harm to users, regardless of the initial intent.

In this paper, we are dedicated to proposing a general framework for evaluating representative recommendation algorithms for the degree of manipulation in user preferences and demonstrating the effectiveness of the proposed framework under various use cases. 

\section{Assumptions}
\label{sec:assumption}
Before presenting our evaluation framework, we need to state two important assumptions. 

First, we expect the proposed framework to work in a fully offline manner to avoid direct interactions with real users. This is because such interactions with evaluation purposes on industrial-level recommender systems are often costly and may harm the platform's revenue. Learning user models from existing offline interaction datasets is an alternative choice. 
In those datasets, we only have access to the observed behavior of the users, such as clicking, but not the reasons behind the behavior. In this case, it is difficult to figure out what the user is really thinking inside their mind. Therefore, we make the following assumption.
\begin{assumption}
\label{assump:user-behavior}
The observed user's behavior is a true reflection of user preferences at that moment.
\end{assumption}
\noindent Notice that the preferences under Assumption \ref{assump:user-behavior} are dynamic, since the clicking behavior of a user at a certain moment may be influenced by items that she has seen before~\cite{pan2022exploiting} or other items on the same slate~\cite{xia2008listwise}.
It is due to this dynamic nature that recommender systems have the potential to increase the platform's revenue by influencing the users' immediate preferences to deviate from their initial preferences prior to interacting with the system. 

As we mentioned in \se{sec:overview-concerns}, the manipulations of users' preferences may be intentional or unintentional. Some might argue that only intentional influences on users' preferences can be called manipulation. We clarify that we do not aim to distinguish the motivations of those manipulations in this study, which is hard and unnecessary from a consequentialist perspective. Therefore, we introduce another assumption to emphasize the simplification.
\begin{assumption}
\label{assump:manipulation}
A recommender system is considered to have manipulated a user's preferences if 1) the user's preferences change during the interaction with the system compared to those before the interaction and 2) this change results in an increase in the platform's revenue, such as clicks or purchases.
\end{assumption}
\noindent With Assumption \ref{assump:manipulation} in mind, as long as we can assess users' initial preferences, their immediate preferences when interacting with a recommender system, and the potential impact of users' behavior on the platform's revenue, we can measure the extent to which that recommender system manipulates users' preferences.

\section{The General Framework for Manipulation Evaluation}
\label{sec:framework}

In this section, we introduce our evaluation framework, \our, in detail. The naming \our implies that we expect to mimic user preference dynamics and reflect the manipulations from recommender systems on user preferences by constructing the simulated user behavior models. Being modular and customizable, this framework is generic and can fit into a wide variety of specific scenarios. We start with the descriptions of the main components involved and then demonstrate the interactive process using these components. The interactive process consists of four stages: 1. initial preference calculation; 2. training data collection; 3. algorithm training and interaction; 4. metrics calculation.
In this section, we illustrate each stage in an abstract manner, and later in experiments we show how each stage can be instantiated according to different scenarios.

\begin{figure}[h]
\centering
\includegraphics[width=0.95\textwidth]{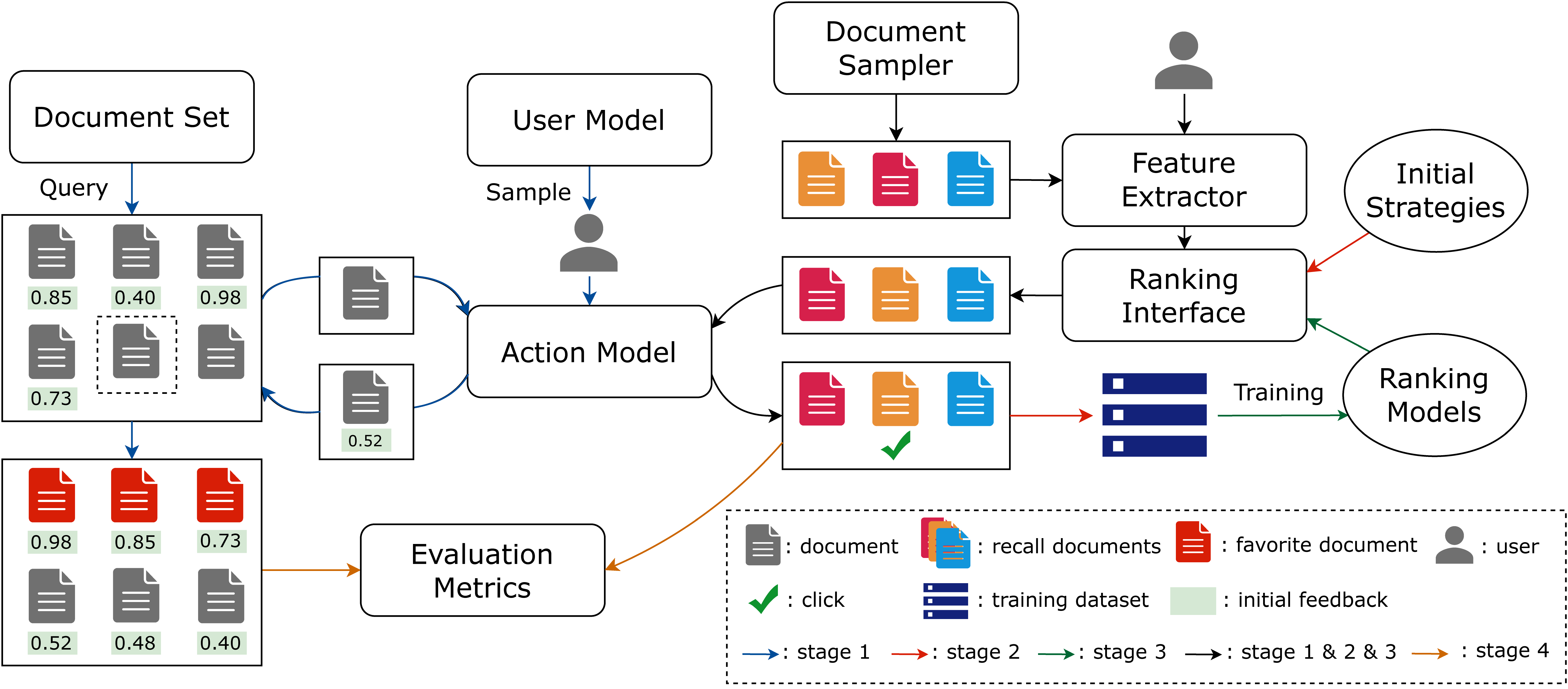}
\caption{The basic components and interactive process of \our.}
\label{fig:benchmark}
\end{figure}

\subsection{Main Components}
\label{sec:main-component}
Here we discuss the main components used in \our. Each of the components can be instantiated with specific implementations, and each combination of these components forms a distinct scenario to evaluate the recommendation algorithms. Among those components, the action model $\mathcal{M}$ is the central part that simulates the users' responses to recommender systems. Different instantiations of the action model can reveal distinct aspects of potential manipulations from recommender systems. We start with other components involved and then elaborate on the details of the action model.

\paragraph{Document Set} The document set $\mathcal{D}$ is a data storage that includes all documents used in the scenario. It should support fetching all available documents under a specific query. Note that in some recommendation scenarios, a user does not start with a query, and any documents can be recommended. In such cases, we assign all users with the same query vector and bind all documents to that single query. 

\paragraph{Document Sampler} In most cases, the complete document set with respect to a query is too large to be directly fed to ranking algorithms. 
Therefore, a document sampler is adopted to recall a subset of all available documents to pass to the ranking algorithms.
We use random sampling or weighted sampling according to historical click-through rates as recall strategies. One can easily integrate more complex recall algorithms into the framework. 

\paragraph{User Model} A set of users $\mathcal{U}$ for recommendation algorithms to interact with are sampled from the user model. A user $i$ is represented by a feature vector $u_i^0$ and a query vector $q_i$. The feature vector not only includes the observable feature that can be used by the recommendation algorithms but can also contain unobservable information, such as the user's current satisfaction level. After viewing a document slate $D$ in interaction round $r$, the user feature changes from $u_i^{r-1}$ to $u_i^r$ following the transition model $\mathcal{T}(u_i^r|u_i^{r-1}, D)$. The query vector is used in the retrieval of a subset of documents from the full document set. 

\paragraph{Feature Extractor} The feature extractor takes as input the document and user feature vector, and outputs the sparse and dense feature for each of them so that ranking models can process each part separately, e.g., use embedding vectors for sparse features. Since the output feature is then passed on to the ranking algorithms, those unobservable features are filtered out here. 

\paragraph{Ranking Interface} The ranking interface aims to provide a unified interface so that researchers can easily integrate new algorithms into the framework and obtain the evaluation results. The framework includes an interaction interface and a training interface. The interaction interface requires the algorithms to provide rankings to a document list when given the observable user features and document features. The training interface is similar to how a general neural network-based recommendation model is trained, where historical interaction data is fed to the algorithm in a batch-wise manner for model updating. The interaction interface is also used by predefined strategies to collect training datasets for model learning. 

\paragraph{Evaluation Metrics} The evaluation metrics are used to describe the benchmark results. Our built-in metrics include those commonly used in recommender systems, such as CTR, NDCG, etc., as well as those needed to evaluate the level of manipulations. These metrics measure the effectiveness of the algorithm while assessing the impact on the user's initial preferences.

\paragraph{Action Model} 
The action model determines how users respond to the recommended items. 
Modules with similar functionality are delicately discussed in other simulation-based work on recommender systems~\cite{zhu2021popularity}, and some of them are referred to by other names, such as interaction model~\cite{chaney2018algorithmic} or user model~\cite{yao2021measuring}. 
Here we do not specify the concrete implementation of the action model, but rather the inputs and outputs that it needs to satisfy. Any function that satisfies the required form, either heuristic rules or data-driven models, can be used as the action model. 
At interaction round $r$, the inputs to the action model are the current user feature $u_i^r$ and recommended document list with $m$ documents $D^r=\{d^r_1,\dots,d^r_m\}$. The outputs are probabilities of different user feedback or responses to each document in the list.
Without loss of generality, we use user clicks as feedback, thus the action model outputs the click probability.
It is easy to incorporate other feedback such as the purchase or like in the framework.
Given the inputs and outputs, the functional form of the action model $\mathcal{M}$ can be written as $p(c^r_{ij}=1)=\mathcal{M}(u_i^{r-1}, d^r_j, D^r_i)$, where $c^r_{ij}=1$ means the user $i$ clicks on the $j$-th item in the recommended list $D_i^r$ in interaction round $r$. 
In real recommendation scenarios, the raw user click probability is not visible to the recommender system. Therefore, the binary user click is generated by sampling from the output probabilities and returned to the recommender system. 

\subsection{Interactive Process}

As long as the main components of a scenario have been specified, the benchmarking of input algorithms can be decomposed as an interactive process consisting of four stages. In the following part, we describe each stage in detail.

\subsubsection{Stage 1: Initial Preference Calculation}

Intuitively, manipulations happen when the users' behaviors are contradictory to their initial preferences. For example, under some carefully designed rankings, the users click on less favored documents with high probabilities compared to the most favored document in the same demonstrated list. Therefore, we have to first access the initial preferences of each user to available documents. 

The definition of initial preferences and the set of related metrics we will later define are central to making \our different from existing simulation-based frameworks, allowing us to quantify manipulations from recommender systems.
We are not ready in this work to propose a definition to measure manipulation from all sources. 
Instead, we study two types of manipulation: manipulation caused by changing the relative order of items within the same slate and manipulation caused by altering the order in which items are presented sequentially. 
These correspond to where recommendation algorithms take effect in two types of recommendation tasks, slate recommendation and sequential recommendation. In order to eliminate these two possible manipulations and obtain the initial preferences, we need to take individual items out of the slate and remove the influence of historical recommendations on the user.

Based on the discussion above, we define the initial preference score $p^0_{ij}$ as the user $i$'s initial feedback to the document $d_j$ when the demonstrated list $\{d_j\}$ only contains one document, i.e., $p^0_{ij}=\mathcal{M}(u_i^0, q_i, d_j, \{d_j\})$. At this stage, we calculate $p^0_{ij}$ for each user $i$ and each document associated with $q_i$, which is determined by the \textit{Document Set}, using the \textit{Action Model}, which is later used for quantifying the deviation of users' preferences. 

\subsubsection{Stage 2: Training Data Collection}

To train a machine-learning recommendation model, we have to build a dataset by interacting with users using some initial recommendation strategies. Note that in real-world cases, the initial strategies are usually previously deployed recommendation models, which are often trained in an offline manner and thus require another batch of interaction data.
To release the burden of such a recursive data-collecting process, we use rule-based strategies or pretrained online algorithms such as reinforcement learning-based methods as the initial data-collection strategy. 
The round $r$ of interaction with a user $i$ starts from retrieving all candidate documents bound to the query $q_i$ from the \textit{Document Set} and then uses the \textit{Document Sampler} to recall a subset of documents $\{d^r_1, \dots, d^r_m\}$. After that, we use the chosen initial strategy to rank the recalled documents as $\{d^r_{k_1}, \dots, d^r_{k_m}\}$, and demonstrate the top-$K$ documents $\{d^r_{k_1}, \dots, d^r_{k_{K}}\}$ to the user, where $K$ is the slate size. The user feedback $c^r_{ij}$ is simulated by the \textit{Action Model} $\mathcal{M}(u_i^{r-1}, d^r_{k_j}, \{d^r_{k_1}, \dots, d^r_{k_{K}}\})$ for $j=1, 2, \dots, K$. The recommended slate, user feature, and feedback are stored in the dataset. In sequential recommendation scenarios, the interactions take place for multiple rounds. 

\begin{figure}[h]
\centering
\includegraphics[width=0.8\textwidth]{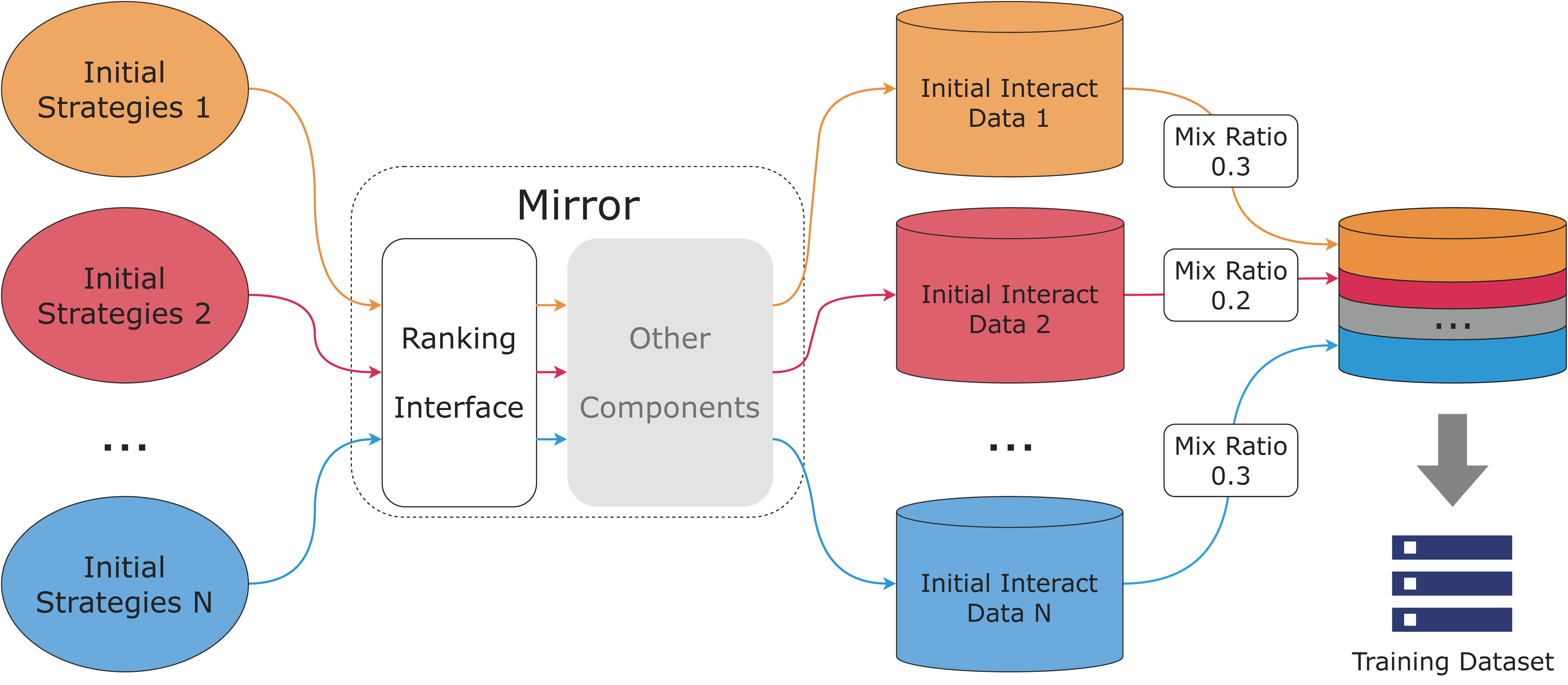}
\caption{An illustration of mixing interaction data from different initial strategies into the training dataset.}
\label{fig:data-process}
\end{figure}

To investigate how certain properties of the training dataset impact the trained algorithms, we need to construct a series of datasets to encompass different values of those properties. For example, suppose we want to study whether more manipulative recommendations in the dataset lead to more manipulative recommendations generated by algorithms trained on it. In that case, we have to obtain datasets with different manipulation levels.
One approach is to design lots of initial strategies with different manipulation degrees, but this can be difficult. 
We take a more tricky approach by first designing two extreme initial strategies, i.e., one that prioritizes based purely on users' initial preferences and one that only considers maximizing user clicks. Intuitively, the latter tends to be highly manipulative. Then we just mix the data collected by these two strategies with different ratios. 
Such mixing of interaction data from different initial strategies can be extended to other properties and mixing of more than two initial strategies. We provide an illustration of data mixing in \fig{fig:data-process}. 

\subsubsection{Stage 3: Algorithm Training and Interaction}

This stage is aimed at simulating the training and deployment of recommendation models in real industry scenarios. Thus we can access the approximate online influences of algorithms made on users by investigating the simulation results. After stage 1, we obtain a group of users with their initial feedback on all related candidate documents. To benchmark a specific algorithm, stage 3 includes training the algorithm on data collected from the users, and interacting with the same group of users using the trained model.

Once the training data is collected and mixed in stage 2, we can train different recommendation algorithms. Depending on the expressive ability of the model and the form of the loss function, the final obtained model will differ in its ability to fit on offline datasets, which will reflect on differences in offline metrics. Traditional evaluations of a recommendation algorithm end up here, without further using the trained model to interact with the users. In our simulation-based framework, we evaluate the online performance of an algorithm by collecting online interacting samples. The interaction procedure is similar to the aforementioned initial data collection, except that the trained model gives the ranking of recalled documents. 

\subsubsection{Stage 4: Metrics Calculation}

In the final stage, after we obtain the interaction data of different algorithms, we can evaluate them with respect to various metrics. Offline metrics can be calculated without interaction data and do not effectively reflect the level of manipulation, so we do not elaborate on them here. We highlight our proposed online metrics, ManiScore defined on a slate, and the Preference Shift (PS) metric used to evaluate the user preference shift during sequential recommendations. Before introducing ManiScore, we first give the definition of the CTR and FCTR on which it depends.

\begin{definition}
Given the user action model $\caM$, the user feature vector $u$, user query vector $q$ and a document list $D=\{d_1, \dots, d_N\}$, define click-through rate (CTR) on the list $D$ as
\begin{equation*}
    CTR=\bbE \Big[\sum_{i=1}^N \frac{c_i}{N}\Big]=\sum_{i=1}^N\frac{\caM(u, q, d_i, D)}{N}~.
\end{equation*}
\end{definition}

As mentioned before, assuming that we have access to the click model $\caM$ and the user's initial preference $u^0$, we can assign each document $d$ in $\caD_q$ with an initial preference score $\phi(d)=\caM(u^0, q, d, \{d\})$, which is the user click probability when the slate only consists of document $d$. Using the initial preference score, we can define the notion of the favorite document set.

\begin{definition}
Given the user action model $\caM$, the initial preference $u^0$, user query vector $q$ and a document set $\caD$, define the top-$K$ favorite document set $\caF^K$ as the set of documents in $\caD$ with $K$ highest initial preference score, i.e., $\caF^K=\argmax_{\caD'\subset\caD, |\caD'|=K}\sum_{d\in\caD'}\caM(u^0, q, d, \{d\})$.
\end{definition}

For a specific dataset, we often use a fixed $K$ and thus abbreviate $\caF^K$ as $\caF$. Now we introduce \textbf{favorite click-through ratio (FCTR)} as a metric to measure how much portion of documents chosen by the user are favorite documents. 

\begin{definition}
Given the user action model $\caM$, the current user feature vector $u$, the user query vector $q$, the favorite document set $\caF$ derived from the complete document set $\caD_q$ and a document list $D=\{d_1, \dots, d_N\}$, define FCTR on list $D$ as
\begin{align*}
    FCTR &= \frac{\sum_{i=1}^N\mathbbm{1}(d_i\in\caF)\caM(u, q, d_i, D)}{\sum_{i=1}^N\caM(u, q, d_i, D)}\\
    &= \bbE_{c_{1:T}\sim\caM}\left[\frac{\sum_{i=1}^N\mathbbm{1}(d_i\in\caF)c_i}{\sum_{i=1}^N c_i}\right]~,
\end{align*}
where $\mathbbm{1}(\cdot)$ is the indicator function.
\end{definition}

\begin{figure}[h]
\centering
\includegraphics[width=0.9\textwidth]{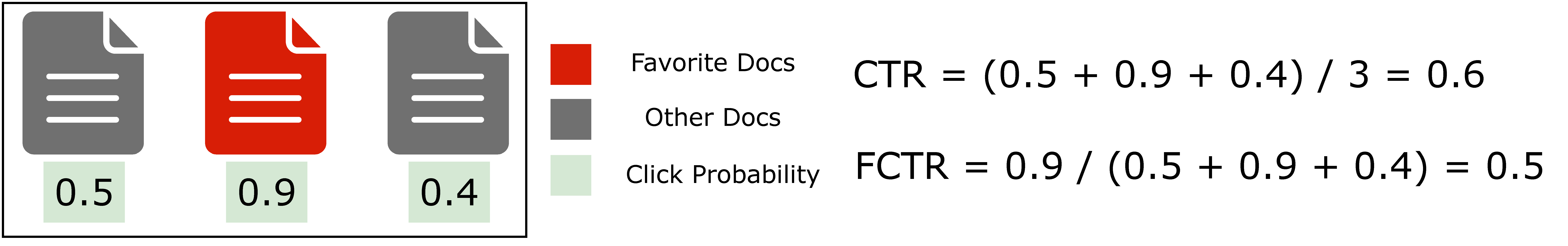}
\caption{CTR and FCTR of an example document list.}
\label{fig:ctr-fctr}
\end{figure}

Intuitively, a lower FCTR means a more significant proportion of user-clicked documents are not initially favored. Note that a lower FCTR does not always relate to a higher CTR, which means such a contradiction of users' preferences may not be effective for improving the platform's income. In this study, we are interested in effective manipulations. To have a comparative anchor for manipulations, we first introduce the definition of unbiased rankings. 

\begin{definition}
\label{def:unbiased-ranking}
Given the user action model $\caM$, the initial user feature vector $u^0$, and a document list $D=\{d_1,\dots,d_N\}$, if a permutation $D'=\{d_{k_1}, \dots, d_{k_N}\}$ of $D$ satisfies $\caM(u^0, d_{k_i}, D')\geq\caM(u^0, d_{k_{i+1}}, D')$ for all $i\geq 1$, then $D'$ is an unbiased ranking of $D$ under $\caM$ and $u^0$.
\end{definition}

We define the \textbf{Manipulation Score (ManiScore)} of a ranked list to take both the relative magnitude CTR and FCTR when compared to unbiased rankings into account.
 
\begin{definition}
Given the user action model $\caM$, the initial user feature vector $u^0$, and a document list $D=\{d_1,\dots,d_N\}$, we denote the set of all unbiased rankings of $D$ under $\caM$ and $u^0$ as $D_U$. We define the Manipulation Score of $D$ as
\begin{equation}
    ManiScore = \exp\Big(\max\Big\{\underbrace{\frac{FCTR_U-FCTR_D}{FCTR_U+FCTR_D}}_{\text{user satisfaction term}}, 0\Big\} + \max\Big\{\underbrace{\frac{CTR_D-CTR_U}{CTR_U+CTR_D}}_{\text{platform revenue term}}, 0\Big\}\Big)~,
    \label{eq:maniscore}
\end{equation}
where $FCTR_U=\min_{D'\in D_U} FCTR_{D'}$ and $CTR_U=\max_{D'\in D_U} CTR_{D'}$.
\end{definition}

Note that ManiScore takes values in $[1, e^2]$. It has a positive correlation with FCTR and a negative correlation with CTR.
Specifically, the user satisfaction term in \eq{eq:maniscore} measures how much the FCTR of the current ranking is lower compared to the unbiased ranking, which indicates whether the recommended list is aligned with the users’ initial preferences.
The platform revenue term measures how much the CTR of the current ranking is higher compared to the unbiased ranking. Since more clicks often bring more revenue to the platform, this term indicates whether platforms benefit from the evaluated recommendation algorithm. The denominators in both terms are to normalize the overall values to a fixed interval. The exponential term is included to make the ManiScore increase slowly when FCTR and CTR do not deviate much from unbiased rankings. 
In particular, a ManiScore of one means the ranked list does not have a lower FCTR nor a higher CTR than any unbiased rankings, which means there are no manipulations. We use ManiScore as the core metric to measure the degree of manipulations of recommender systems.

During sequential recommendation, we introduce the \textbf{PS (Preference Shift)} metric to evaluate user preference shift across different rounds.

\begin{definition}
Given the user action model $\caM$, the user preference $u^r$, and a document set $\caD$, define the top-$K$ favorite document set of interaction round $r$, $\caF^K_r$ as the set of documents in $\caD$ with $K$ highest preference score in round $r$, i.e., $\caF^K_r=\argmax_{\caD'\subset\caD, |\caD'|=K}\sum_{d\in\caD'}\caM(u^r, d, \{d\})$.
We further sort the document in $\caF^K_r$ in descending order according to preference score $\caM(u^r, d, \{d\})$ and get $\caL^K_r$.
When $r=0$, $\caL^K_r$ is similar to $\caF^K$ in Definition 3.2, except for the sorting process.
Obtaining the sequence of favorite document list in different interact rounds, $\caL_0, \caL_1, \ldots \caL_n$, we now define preference shift of round $r$ as
\begin{equation*}
\begin{aligned}
PS_r
&= 1 - Similarity(\caL_0, \caL_r)= 1 - RBO(\caL_0, \caL_r, p) \\
&= 1 - (1-p)\sum_{d=1}^{\infty} p^{d-1} \cdot \frac{\mid\caL_{0, :d}\cap\caL_{r, :d} \mid}{d}~,
\end{aligned}
\end{equation*}
where $d$ stands for the depth of the ranking list $\caL_0, \caL_r$, and $p$ is a parameter that determines how steep the decline in weights is: the smaller $p$, the more top-weighted the metric is.
\end{definition}

Note that rank-biased overlap~\cite{webber2010similarity} (RBO) is a measure that calculates the similarity between incomplete rankings, which could handle non-conjointness.
As a result, preference shift falls in the range [0, 1], where zero means identical to the initial preference, and one means totally different.

\section{Experiment}
\label{sec:exp}

In this section, we employ our proposed benchmarking framework on both slate and sequential recommendation scenarios to investigate the manipulations of different algorithms on users' preferences. 
Specifically, we conduct two synthetic experiments based on users' behavioral analysis and two real-world experiments based on existing data-driven simulation environments~\cite{dai2021adversarial, wang2021rl4rs}. Individual conclusions are made after the presentation of the numerical results of each experiment. We conclude this section with a comprehensive analysis of the effects of all four experiments, in which we state our core findings.


\subsection{Synthetic Slate Recommendation Experiment}

\subsubsection{Experiment Settings}
Consider the scenario where the user has to choose from multiple forms of transportation. Each form of transportation is evaluated by the user for its traveling time $h$ and price $m$. We assume the probability of the user's choice of transportation $d_i$ follows a generalized Random Regret Minimization (RRM) model~\cite{chorus2014random}:
\begin{equation*}
    p_{\text{RRM}}(c_i=1|D)=\frac{\exp(-RR_i)}{\exp(L)+\sum_j\exp(-RR_j)}~,
\end{equation*}
where $D$ is the set of all transportation forms examined by the user, and $L$ is the value to model the probability of not choosing any displayed transportation. Here we set $L=-1$. The random regret $RR$ is given by
\begin{equation*}
    RR_i=R_i+\epsilon_i=\sum_{d_j\neq d_i, d_j\in D}\left[\max\{h_i-h_j, 0\}+\max\{m_i-m_j, 0\}\right]+\epsilon_i~.
\end{equation*}

For simplicity, we set all $\epsilon_i$'s to zero. Note that the RRM model only takes the examined documents into account. Different documents will have the same $R$ under the RRM model when calculating the initial preferences since there is always no regret when a slate only contains one document. 
Therefore, we augment the RRM action model by defining the user click probability when only one transportation $d$ is examined as
\begin{equation*}
    p_{\text{RRM}}(c=1|\{d\})=\frac{\exp(U)}{\exp(L)+\exp(U)}~,
\end{equation*}
where the utility $U$ is given by
\begin{equation*}
    U=-(h+m)~.
\end{equation*}

In this synthetic experiment, each slate demonstrated to the user consists of three items. To fit in the slate recommendation scenario and consider the user's view order, we set the examination probability of each three positions in the slate as $[1.0, 0.8, 0.6]$.
Therefore, for a recommended slate $D=\{d_1, d_2, d_3\}$, the click probability for $d_i$ is
\begin{equation*}
    p(c_i=1)=\sum_{e_1, e_2, e_3\in\{0, 1\}} (1.0)^{e_1} (0.8)^{e_2} (0.6)^{e_3} \mathbbm{1}(e_i=1) p_{\text{RRM}}(c_i=1|\{d_i:e_i=1\})~,
\end{equation*}
where $\mathbbm{1}(\cdot)$ is the indicator function.

Intuitively, the transportation with less traveling time and a lower price has higher initial preference scores. The RRM model is shown to have the potential to induce the decoy effect~\cite{guevara2016modeling}.
Decoy effect~\cite{huber1982adding, min2003consumer} refers to the following phenomenon observed in human decision-making. 
Suppose there are two items for the user or consumer to choose from, and the first item is of higher quality and, at the same time, more expensive than the second item. In this situation, the user needs to make a trade-off between quality and price.
However, if a third item is added with its price between those two items and its quality being the lowest, the user's choice will be influenced. Now the second item is better than the newly added one in terms of both price and quality, and the user will have a higher chance to choose the second item. 
The third item is not added to be chosen by the user, but as a reference for comparison to make the user prefer the second item among the first two. Therefore, the third item is also called a decoy. 
The case that needs to be highlighted is that the first item is a better choice when considering both aspects comprehensively. Then, the introduction of the decoy can lead to irrational behavior of the user.
Note that in the case of the transportation scenario, the faster transportation form can be viewed as having higher quality.

Under our framework, we aim to use a synthetic dataset to study the relationship between manipulations of users' preferences and the underlying decoy effect, and how much different ranking algorithms can use the decoy effect to increase traditional online metrics. 
We start with a numerical example to further illustrate the decoy effect in the transportation scenario. 
Now we consider these four choices of transportation: $d_1$ with $h_1=0.8$ and $m_1=0.2$, $d_2$ with $h_2=1.0$ and $m_2=0.4$, $d_3$ with $h_3=0.1$ and $m_3=0.8$, and $d_4$ with $h_4=0.45$ and $m_4=0.5$. 
We define the user's favorite document as the one with the highest initial preference score, i.e., $d_3$. Compared to $d_3$, $d_1$ has a lower initial preference score, but $d_2$ can serve as a decoy of $d_1$. 
We can see that $d_2$ is similar to $d_1$ but is worse in both aspects, which satisfies the requirements in the decoy effect. During each round of interaction, the document sampler will recall one of two lists of documents, either $d_1, d_2, d_3$ or $d_1, d_3, d_4$. 

\begin{figure}[htbp]
    \centering
    \includegraphics[width=0.8\textwidth]{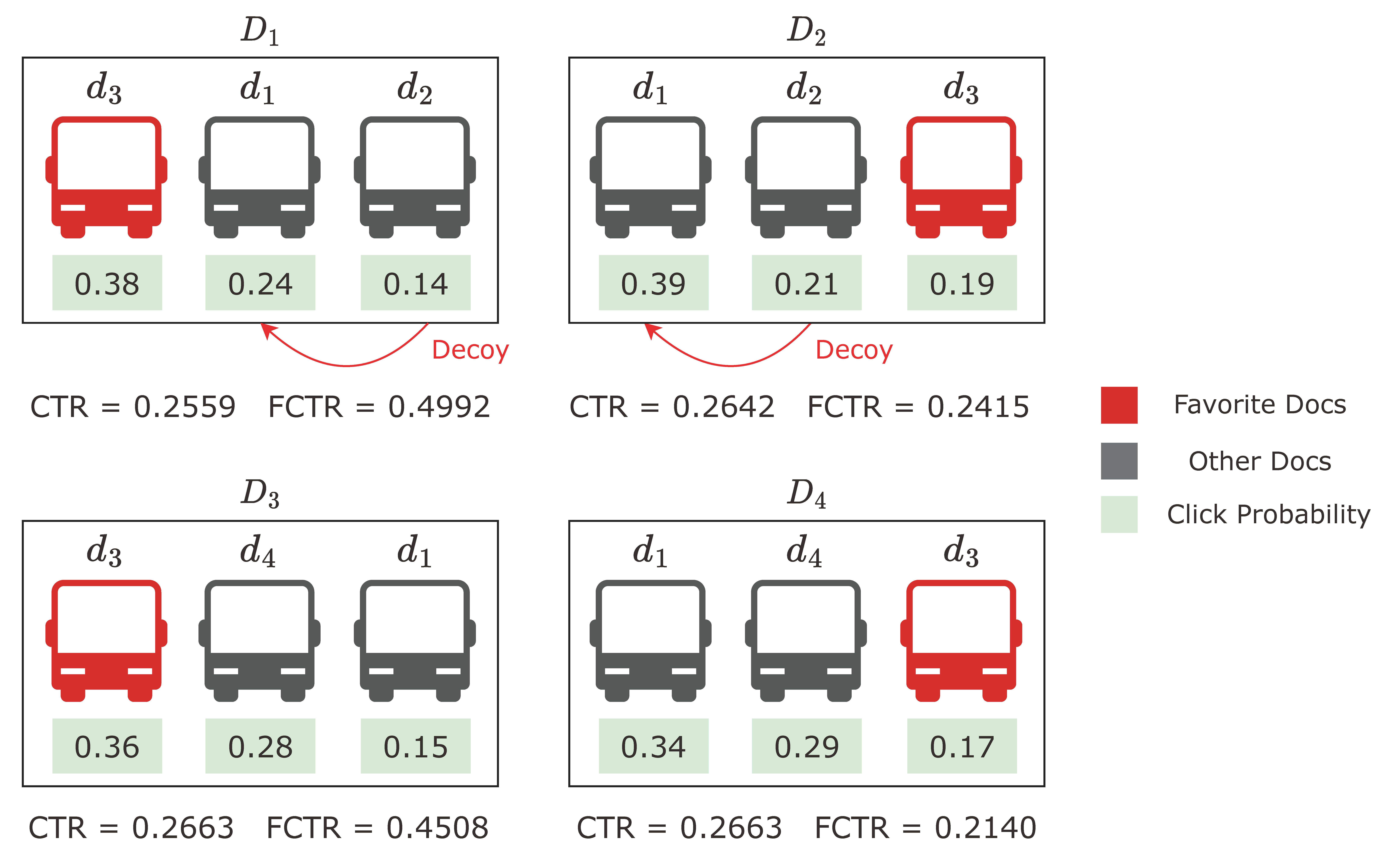}
    \caption{User feedback and metrics of different rankings for two recall lists under the RRM action model.}
    \label{fig:decoy-slates}
\end{figure}

In \fig{fig:decoy-slates}, we demonstrate two different ranking lists for each recall list, their corresponding user click probabilities, and online metrics (CTR and FCTR). Red arrows visualize the decoy effect. When comparing $D_1$ and $D_2$, we can see that placing the pair of documents with the decoy effect at the top of the slate, $D_2$ has higher CTR and lower FCTR. On the other hand, for $D_3$ and $D_4$ as rankings for the other recall list, $d_1$ and $d_3$ are paired with the `medium' document $d_4$, which does not have a decoy effect on either of them. In this case, when placed at the same position, $d_3$ receives a higher click probability compared to $d_1$ for its advantage.

In practical implementations, we introduce randomness in the document properties to have a richer and more diverse document set. The document set is made up of 100 groups of documents, and each group consists of four documents from $d_1$ to $d_4$. The traveling time and price in the first 50 document groups are sampled or set by the following rules:
\begin{equation*}
\begin{aligned}
h_1\sim 0.5+\text{Uniform}(0, 0.5), & \quad m_1=1-h_1\\
h_2\sim h_1+\text{Uniform}(0, 0.1), & \quad m_2\sim m_1+\text{Uniform}(0, 0.1)\\
h_3\sim \max\{1 - h_1 - \text{Uniform}(0, 0.1), 0\}, & \quad m_3\sim\max\{1-m_1-\text{Uniform}(0, 0.1), 0\}\\
h_4=(h_1+h_3)/2, & \quad m_4=(m_1+m_3)/2~,
\end{aligned}
\end{equation*}
where $\text{Uniform}(a, b)$ is the uniform distribution defined in $[a, b]$. In most cases, the four documents in each group obtained in such a manner can produce the decoy effect described in \fig{fig:decoy-slates}.
The distributions of traveling time and price dimensions are not aligned by such a sampling procedure, which makes it unreasonable to add them two together when calculating utilities. To address the issue, the remaining 50 groups of documents are generated by swapping the numerical values of the traveling time and price of those documents in the first 50 groups.

Here we only introduce diversity in document attributes and do not consider the difference in preferences between different users. That is, this synthetic experiment can be seen as evaluating the interaction results of multiple recommendation algorithms on a single user. 
The reason is that in this part, we focus on investigating whether different recommendation algorithms can manipulate users with explicitly designed non-optimality. 
Such manipulations do not necessarily rely on modeling similarities between users and can be learned in the interaction data with a single non-optimal user. 
To keep the synthetic experiments concise, we do not add differences in user preferences.

To generate training data with different degrees of manipulation, we use the following two strategies to collect the initial dataset. 
\begin{itemize}
    \item \textbf{Decoy Oracle} is the strategy that ranks $d_1, d_2, d_4$ as $D_2$ and ranks $d_1, d_3, d_4$ as $D_3$ and $D_4$ for equal probability. 
    \item \textbf{Unbiased Oracle} is the strategy that ranks $d_1, d_2, d_4$ as $D_1$ and ranks $d_1, d_3, d_4$ as $D_3$ and $D_4$ for equal probability. 
\end{itemize}

We generate a training dataset by mixing up the interaction data collected by these two ranking strategies with different mix ratios. A mix ratio of $\alpha$ means $\alpha$ fraction of training data are ranked by Decoy Oracle and the rest $(1-\alpha)$ are ranked by Unbiased Oracle. 

\subsubsection{Evaluated Methods}
We evaluate two LTR algorithms in this scenario. 

\begin{itemize}
    \item \textbf{DeepFM} \cite{guo2017deepfm} is a neural ranking model that combines a factorization machine and a deep neural network. It is a representative work of point-wise ranking models that treat each item independently. 
    \item \textbf{SetRank} \cite{pang2020setrank} is a neural reranking algorithm using a self-attention structure to encode the entire slate and is trained with attention loss. We use SetRank as an instance to represent reranking algorithms, which can model the mutual interactions between items in the same slate. 
\end{itemize}

\subsubsection{Experiment Results}


\begin{figure}[h]
    \centering
    \includegraphics[width=0.9\textwidth]{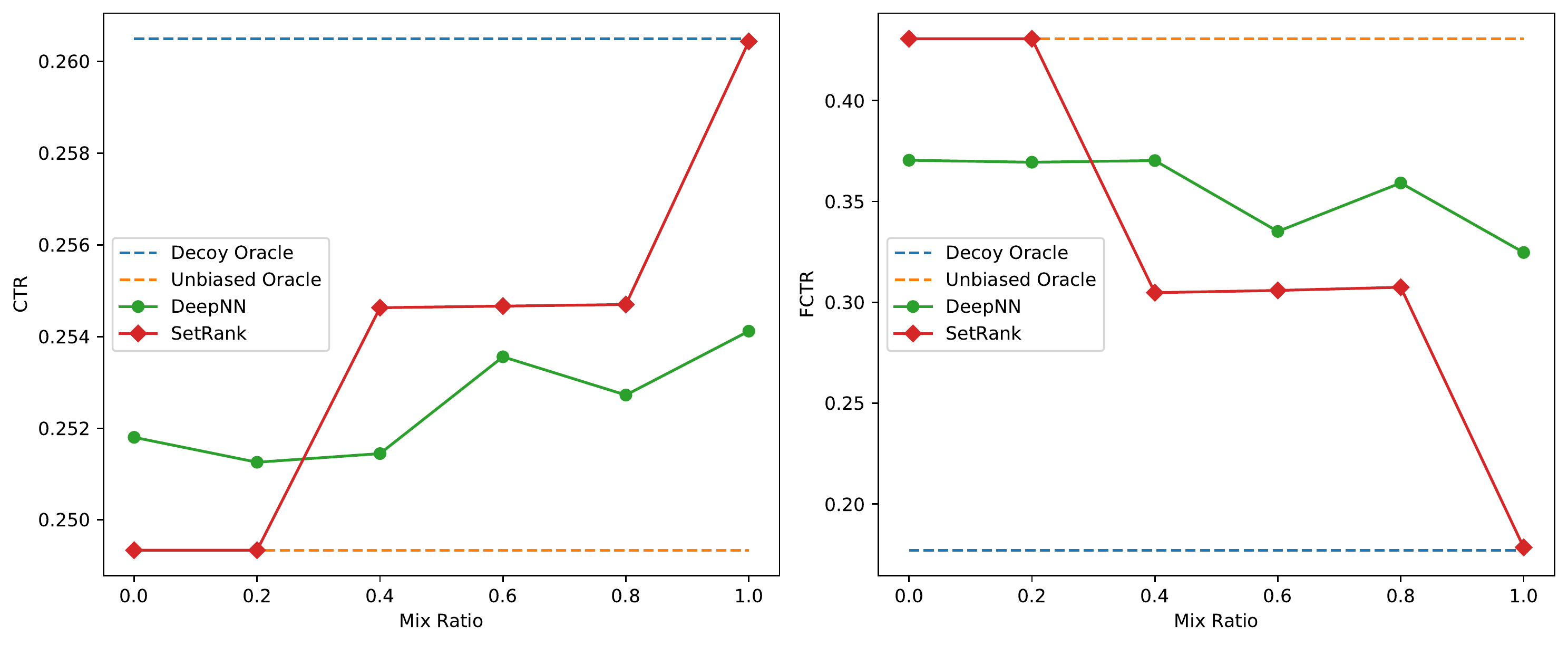}
    \caption{The CTR and FCTR of algorithms trained on the synthetic dataset with different mix ratios. The metrics values of initial strategies are plotted with dashed lines.}
    \label{fig:decoy-mix-ratio}
\end{figure}

From \fig{fig:decoy-mix-ratio}, we can see that when the mix ratio of training data is 1.0, ranking lists given by SetRank have higher CTR and lower FCTR than DeepFM. When the mix ratio increases from 0.0 to 1.0, both metrics of DeepFM change smoothly, while those of SetRank experience a sudden change somewhere around the mix ratio of 0.4 and 0.8. 
In general, as more data generated from Decoy Oracle are mixed into the dataset, the CTR of both methods becomes higher while their FCTR becomes lower. 

\subsubsection{Conclusions}
From this experiment, we can draw the following conclusions:  1) the degree of manipulation is mainly influenced by the ranking strategy used to generate the training data; 2) When there exist manipulations in the training data, the reranking algorithms like SetRank are more likely to learn from those patterns of manipulation compared with point-wise methods like DeepFM. It is reasonable since reranking algorithms can model the mutual influences among the entire slate and thus are better aware of the decoy effects in the slate.

\subsection{Data-driven Slate Recommendation Experiment}

\subsubsection{Experiment Settings}

In this subsection, we conduct data-driven experiments under the slate recommendation setting. We use AICM~\cite{dai2021adversarial} with TianGong-ST~\cite{chen2019tiangong} dataset as our test bed. AICM adopts a GAN-like adversarial learning approach to imitate human users' click behavior. A generator is used as the simulated user click model, and a discriminator measures the distance between generated clicks and human clicks in the dataset. Experiments show that the generator in AICM can generalize well across different ranked list distributions when compared with other neural click models, including NCM~\cite{borisov2016neural} and CACM~\cite{chen2020context}. 
Such superior distributional coverage makes the generator quite suitable to be used as simulated human users when interacting with diverse recommendation models. 

Before using the trained generator in AICM as an action model in \our, we need to verify that it meets the input and output forms required in \se{sec:main-component}.
The generator is implemented using the gated recurrent unit (GRU)~\cite{cho2014properties} structure.
As slate recommendations only involve a single round of interaction, we omit the superscript $r$ for simplicity.
For each document list $D=\{d_1, \dots, d_m\}$, the generator processes each document $d_j$ in a top-to-bottom order, which allows the documents that are ranked on top to influence the click probabilities of the documents that follow. The current user feature $u_i$, document feature $d_j$ and the click on the previous document $c_{i(j-1)}$ are used to update the hidden state of GRU from $h_{i(j-1)}$ to $h_{ij}$. Then a linear layer followed by a Softmax operator transforms the updated hidden state to the click probability $p(c_{ij}=1)$. Therefore, the computation inside the generator can be written as
\begin{align*}
    [p(c_{ij}=1), p(c_{ij}=0)]&=\text{Softmax}(\text{Linear}(h_{ij})) \\
    &=\text{Softmax}(\text{Linear}(\text{GRU}(h_{i(j-1)}, u_i, d_j)))~.
\end{align*}
We know that $h_{i(j-1)}$ is updated by all previous documents $\{d_1, \dots, d_{j-1}\}$ and $u_i$. Thus, the input used to compute the output probability $p(c_{ij}=1)$ includes $\{d_1, \dots, d_{j}\}$ and $u_i$, which is a subset of the required form of input. That is, the generator in AICM meets the requirements of the action model in \our.

In our experiment, we employ 35,000 queries from the TianGong-ST dataset to form our user set, and each query is associated with ten candidate documents. Since we are interested in the manipulations that happen in real-world recommendation systems, we use the same set of users for training data collection and online evaluation. The original page size of AICM is ten, and here we use five as the page size to allow randomness among interactions with the same user. During each interaction, half of the documents are randomly recalled from ten candidates and then ranked by the scores provided by the ranking or reranking algorithm. To collect offline data with different degrees of manipulations, we introduce two recommendation policies that represent the most and least manipulations possible, respectively.
\begin{itemize}
    \item \textbf{Greedy Oracle} has access to the ground-truth user model and will enumerate all permutations of items on the slate and select the one with the highest click probability.
    \item \textbf{Unbiased Oracle} is the strategy that consistently ranks items according to users' initial preference scores. For a given document set $D$, it provides one of the unbiased rankings, as in \defi{def:unbiased-ranking}.
\end{itemize}
We construct offline data with different degrees of manipulation by mixing up the user feedback of slates generated by these two oracles with different ratios. Specifically, the offline data consists of feedback for ten pages from each user, and the mix ratio is the percentage of the slates generated from Greedy Oracle. For example, a mix ratio of 1.0 means all data are generated from Greedy Oracle, 0.5 means for each user, Unbiased Oracle ranks five slates, and Greedy Oracle ranks the other five. When training all the algorithms, we use $15\%$ of the offline data as the test set and the rest as the training set.

\subsubsection{Evaluated Methods}
Besides two LTR algorithms already introduced in the synthetic experiment, we include one more point-wise ranking algorithm with different loss functions in this experiment.
\begin{itemize}
    \item \textbf{LambdaFM} \cite{yuan2016lambdafm} shares the same encoding module with DeepFM but is trained with pair-wise loss functions weighted by delta NDCG to reduce the incorrectly ordered pairs. It is a typical example of a point-wise model trained with pair-wise loss functions.
\end{itemize}

\subsubsection{Experiment Results}

First, we present offline and online metrics of the evaluated algorithms trained on the same data with a mixed ratio of 0.5. Before the analysis of the learning algorithms, we can see there exists a vast gap between the CTR and FCTR of two oracles, which means the significant impact of the ranking of documents in the recommended list on user behaviors. As we can see from \tb{tb:slate-result}, with modeling capabilities at the list level, SetRank obtains the best performance in offline metrics and online CTR compared to other learning algorithms. The simple point-wise model, DeepFM, performs the worst in these metrics.
However, the FCTR of SetRank is significantly lower than that of DeepFM, which means the high CTR of SetRank is accompanied by users clicking even more items they do not favor. We also notice that LambdaFM has both lower CTR and FCTR compared to SetRank, which indicates that not all infringements of users' preferences are effective manipulations. 

\begin{center}
\begin{table}[h]
\begin{tabular}{c|cc|ccc}
    \hline
    \hline
    \multirow{2}*{Method}& \multicolumn{2}{c|}{Offline} & \multicolumn{3}{c}{Online}\\
    ~ & NDCG & AUC & CTR & FCTR & ManiScore\\
    \hline
    DeepFM & 0.5685 & 0.6602 & 0.2010 & \textbf{0.7467} & \textbf{1.1217} \\
    LambdaFM & 0.5749 & 0.6669 & 0.2012 & 0.7188 & 1.1537 \\
    SetRank & \textbf{0.5835}  & \textbf{0.7054} & \textbf{0.2031} & 0.7293 & 1.1382 \\
    \hline
    Unbiased Oracle & 0.5884 & - & 0.1961 & 0.8383 & 1\\
    Greedy Oracle & 0.5476 & - & 0.2323& 0.5500 & 1.3838\\
    \hline
    \hline
\end{tabular}
\vspace{5pt}
\caption{The evaluation results of different algorithms (strategies) on the data-driven slate recommendation scenario, where the mix ratio of training data is 0.5.}
\label{tb:slate-result}
\end{table}
\vspace{-10pt}
\end{center}

\begin{figure}[htbp]
    \centering
    \subfigure[]{
        \includegraphics[scale=0.33]{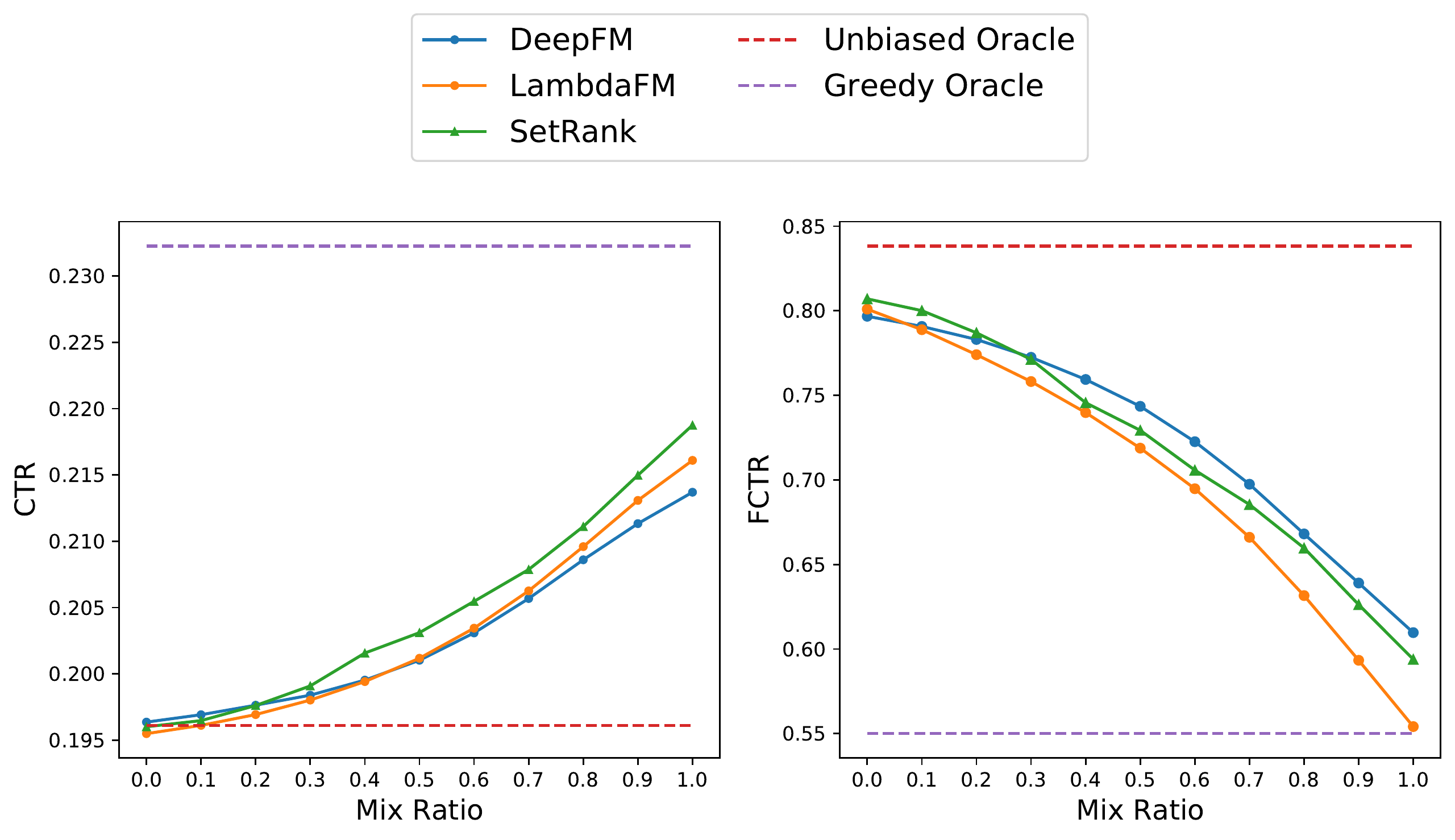}
        \label{fig:slate-mix-line}
    }
    \subfigure[]{
	    \includegraphics[scale=0.33]{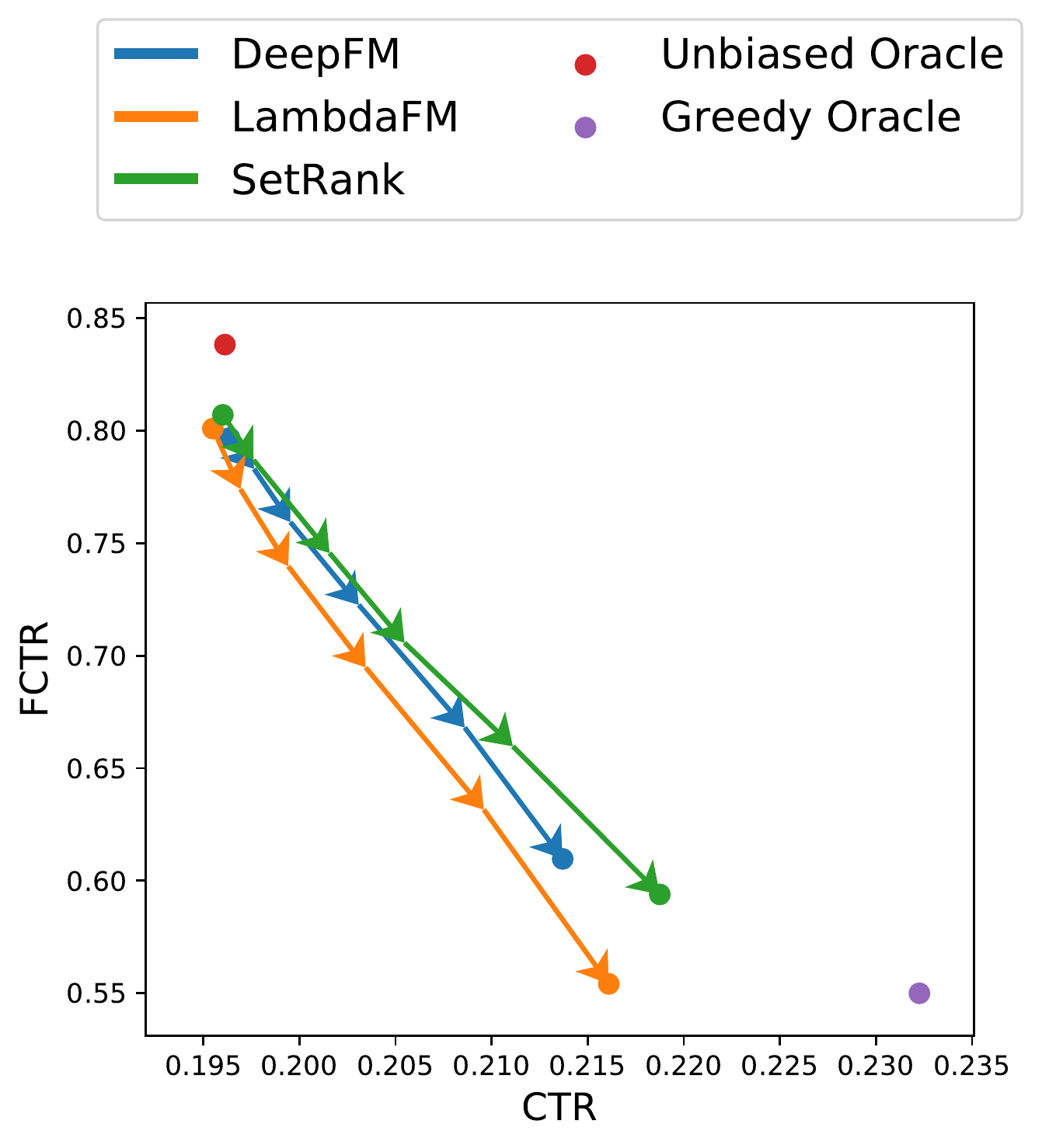}
	    \label{fig:slate-mix-scatter}
    }
    \caption{Online metrics of different recommendation algorithms when trained with data of different mix ratios. Figure (a) shows the changing curve of CTR and FCTR in the same figure, where the arrows point from the low mix ratios to the high mix ratios.
    Figure (b) shows the changing curve of CTR and FCTR in the same figure, where the arrows point from the low mix ratios to the high mix ratios.}
\end{figure}

We also train those three recommendation algorithms using training data with different mix ratios. The changing curve of CTR and FCTR with various mix ratios are plotted in \fig{fig:slate-mix-line}. Overall, as the mixed ratio increases, i.e., more training data is generated by Greedy Oracle, all algorithms tend to have lower FCTR and higher CTR. It means that the degree of manipulation of the learned algorithm is easily influenced by the degree of manipulation of the recommendation strategy in the training data. As illustrated in \fig{fig:slate-mix-scatter}, if we connect the corresponding points of the same algorithm from lower to higher mix ratios to form a trajectory on CTR-FCTR space, the trend of coming close to the training data point is quite clear. Another interesting finding is that when the mix ratio is very close to 1.0, SetRank instead has the highest FCTR. It indicates that SetRank can better fit the unbiased data for its stronger modeling capacity when it is clean of manipulations.

\subsubsection{Case Study}

\begin{figure}[h]
\centering
\includegraphics[width=1.0\textwidth]{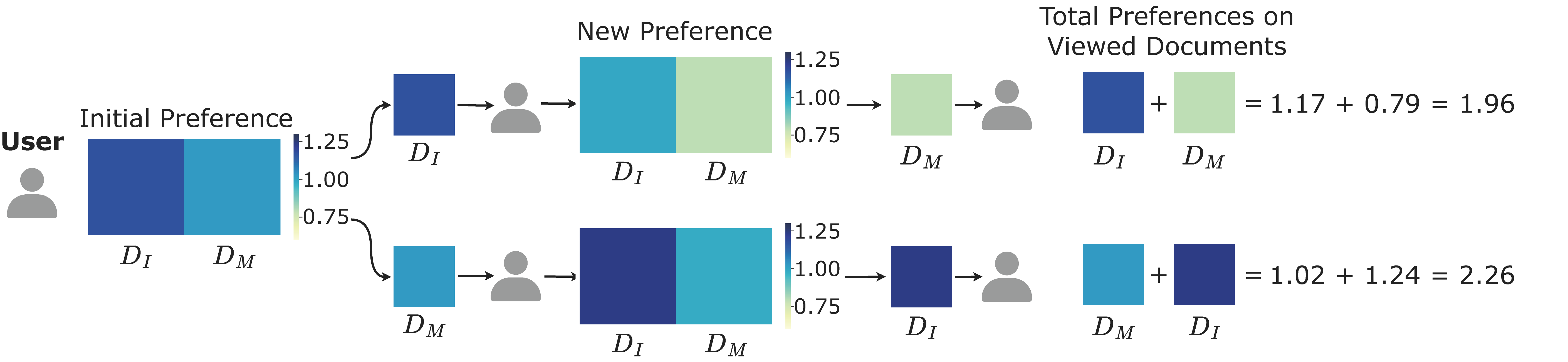}
\caption{Comparison of the unbiased and greedy data in user model.}
\label{fig:preference}
\end{figure}

We take an example from our dataset to better understand the unbiased and greedy policy.
We first take a glance at the action model.
For each user, we sort the candidate items according to their initial preference and choose two of them, where one item $d_{I}$ is of the highest initial preference to the user, implying that it is more likely to be ranked first by an unbiased policy. The other item $d_M$ is of medium preference to the user, suggesting that it is more likely to be ranked first by a greedy policy.
In Fig.~\ref{fig:preference}, we show the difference between unbiased and greedy policies. The first row stands for an unbiased policy, and the second stands for a greedy policy, where the unbiased policy first shows $d_I$ to the user, and the greedy policy first shows $d_M$.
The user will re-evaluate the two items when the first item is shown to the user. 
For the unbiased policy, the first viewed item by the user is of the highest initial preference, so the user tends to give a lower relevant score during re-evaluations. 
In contrast, for the greedy policy, as the first item is of medium preference, a higher relevant score will be given to the two items after the first item is shown.
When the two items are shown to the user successively, we calculate the overall user preference towards the same two items and find that the greedy policy gains a higher score.

Now we have seen there do exist unbiased and greedy data in the user model, and the next question is the capability of different algorithms to manipulate.
We fix the mix ratio of training data as zero, which means that the training data is purely greedy.
In Table~\ref{tb:diff_level_manipulate}, the average position of the favorite item means the average position of the users' favorite item that occurs in the slate (which ranges from 1 to 5), where the favorite item means the user has the highest initial preference for the item. Similarly, we can define the average position of the least favorite item. 
Intuitively, the higher the degree of manipulation is,
the larger the average position of the favorite item is, and the smaller the average position of the least favorite item is.
Table~\ref{tb:diff_level_manipulate} shows that SetRank manipulates more than DeepFM when the training data is manipulated.

\begin{center}
\begin{table}[h]
\begin{tabular}{c|c|c}
    \hline
    \hline
    {Method}& 
    \makecell[c]{Average Position \\ of the Favorite Item} & 
    \makecell[c]{Average Position \\ of the Least Favorite Item}\\
    \hline
    DeepFM & 2.5621 & 3.4738\\
    SetRank & 2.9990  & 2.9964 \\
    \hline
    \hline
\end{tabular}
\vspace{5pt}
\caption{The average positions of both types of documents in the recommended slates generated by different algorithms.}
\label{tb:diff_level_manipulate}
\vspace{-20pt}
\end{table}
\end{center}

\subsubsection{Conclusions}

From this experiment, we can draw the following conclusions: 1) real users can also be heavily manipulated by the recommendation strategy, sacrificing FCTR to achieve a higher CTR; 2) the findings in the synthetic experiment regarding the effects of training data and model choice on the degree of manipulations still apply to real users; 3) the case study reveals that manipulations on real users can be explained by placing less favored documents in front of the favorite items, similar to the decoy effect in the synthetic experiment.






\subsection{Synthetic Sequential Recommendation Experiment}

\subsubsection{Experiment Settings}

We start the evaluation of sequential recommendation algorithms with a synthetic scenario adopted by previous works~\cite{ie2019recsim, ie2019slateq}. In this scenario, each document in the document set belongs to a topic, where in total, we have $m$ topics. The user feature vector $u\in[0, 1]^m$ is the users' preference for each topic. In each round of sequential interaction, the recommendation algorithm presents users with a document from a set of candidate documents. For each document, besides the topic represented by a one-hot vector, it is also assigned with an unobservable scalar quality. The quality is sampled from Gaussian distributions with fixed variance and means determined by the topic. Initially, users have a fixed time budget, and the time budget will reduce after each round. The quality of the document specifies the reduced amount. If the document quality is high, the number of total viewed documents will be higher. Therefore, this leads to a balance between long-term and short-term gains. The recommendation algorithm can change users' preferences by influencing them in the long term so that they all move to higher-quality documents on average to achieve more clicks. 

In our experiment, we set $m=10$, and the document set consists of 10 documents for each topic, which in total has 100 documents. The algorithm can interact with users for at most 20 rounds. During each round, the ranking algorithms select 1 document from 10 randomly recalled documents (from 100 documents) to demonstrate to the user.

We use two strategies to generate a diverse training dataset:
\begin{itemize}
    \item \textbf{Unbiased Oracle} always selects recommended documents based on the user's initial interest.
    \item \textbf{Planner Oracle} evaluates a document by computing the expected cumulative clicks when recommending documents from the same topic in each subsequent round using the mean quality of that topic and makes recommendations based on the evaluation.
\end{itemize}
Intuitively, the Planner strategy may recommend documents from topics not currently the user's favorite in the short term, thus getting more clicks in the long term. The training data is made up of half of the data collected using Unbiased Oracle and the other half collected using Planner Oracle. 

\subsubsection{Evaluated Methods}

In this scenario, we adopt two ranking algorithms for comparison.
\begin{itemize}
    \item \textbf{DNN} uses a feedforward neural network to encode the user feature and document features and does not model the user history. The network is trained with cross-entropy loss to predict the user click probability. 
    \item \textbf{GRU} builds upon DNN and additionally uses a GRU network to model clicked documents during the current interaction round. We use this algorithm as a representative of sequential recommendation algorithms.
\end{itemize}

\subsubsection{Experimental Results}

\begin{center}
\begin{table}[h]
\begin{tabular}{c|cccc}
    \hline
    \hline
    Method & Offline AUC & CTR & FCTR & PS \\
    \hline
    DNN & 0.6193 & 0.3725 & \textbf{0.5283} & \textbf{0.0074} \\
    GRU & \textbf{0.6226} & \textbf{0.3827} & 0.3195 & 0.0140 \\
    \hline
    Unbiased Oracle & - & 0.3633 & 0.6692 & 0.0024  \\
    Planner Oracle & - & 0.3860 & 0.2801 & 0.0203 \\
    \hline
    \hline
\end{tabular}
\vspace{5pt}
\caption{The evaluation results of different algorithms (strategies) on the synthetic sequential scenario.}
\label{tb:iev-result}
\end{table}
\end{center}

As we can see in \tb{tb:iev-result}, the two oracle strategies perform as expected. Comparing the evaluation results of the two strategies, the Planner Oracle has a higher CTR and a lower FCTR. Moreover, we demonstrate the users' Preference Shift after 20 rounds of interaction using both strategies. The Planner Oracle has an order of magnitude higher impact on user preferences than the Unbiased Oracle. Notice that the PS of both strategies is low since, in this scenario, there are ten topics of documents in total, and only the preferences of the topics that are demonstrated to the users change, the relative relationships of most topics remain the same. For the two algorithms evaluated, when trained with the same dataset, GRU behaves more closely to Planner in online metrics than DNN. GRU obtains a higher CTR at the cost of a lower FCTR and more significant impacts on user preferences. On the other hand, GRU has a higher offline AUC, which means it fits better to the dataset and is superior considering pure offline evaluations.

\begin{figure}[h]
\centering
\includegraphics[width=0.9\textwidth]{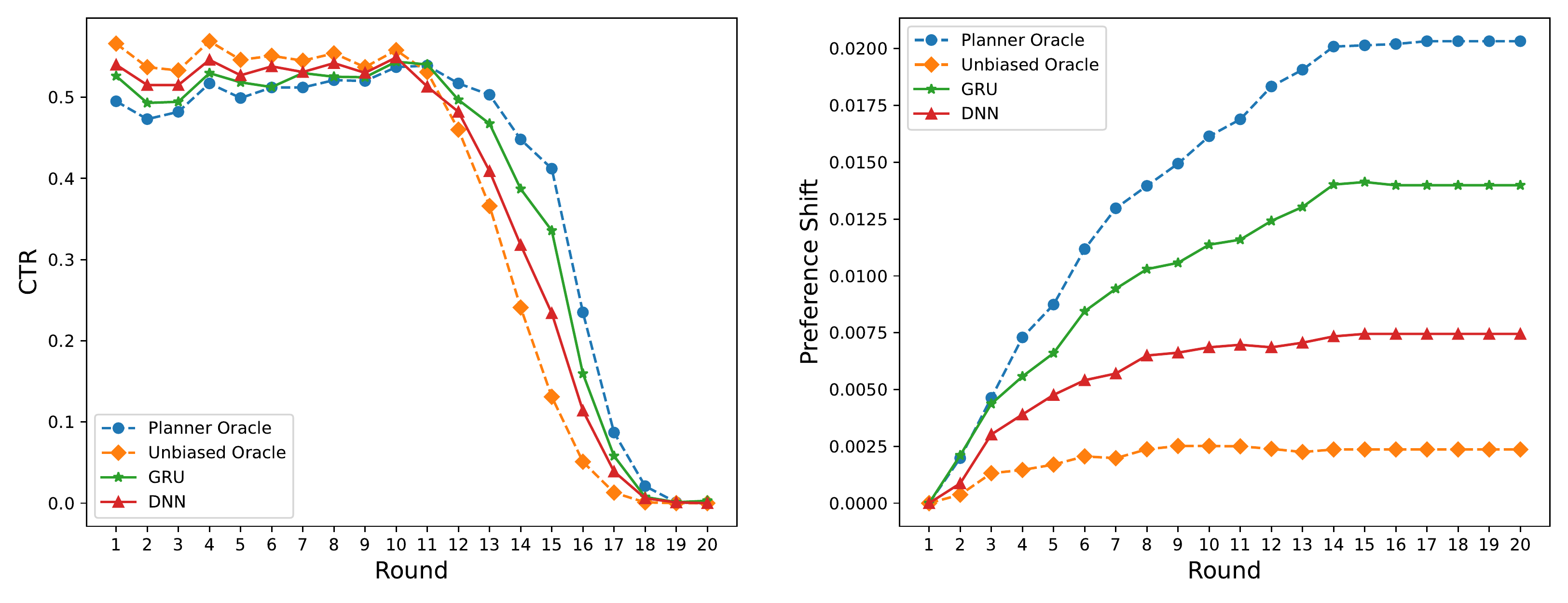}
\Description{}
\caption{CTR and PS of evaluated algorithms and strategies across 20 interact rounds.}
\label{fig:iev-result}
\end{figure}

We also show in \fig{fig:iev-result} the trend of CTR and PS with the number of interaction rounds in the sequential recommendation. Note that when we calculate CTR, users who run out of time budget are included in the calculation as samples with no clicks, so as the number of rounds increases and more users leave the platform, the overall CTR will tend to decrease. Comparing the two Oracle strategies, we can see that Planner Oracle has a lower CTR than Unbiased Oracle in the first nine rounds and a higher CTR than Unbiased from 10 rounds onwards, which is due to the slower decline of Planner Oracle's CTR. It can be explained by Planner Oracle's tendency to recommend documents that will entice users to stay on the platform longer rather than what they are most likely to click on at the moment. Documents within a user's current preferred topic may be of lower quality, and the Planner will actively change the user's preferences to achieve higher CTR throughout the interaction rounds. The higher PS corresponding to the Planner Oracle can also account for its more active influence on user preferences. Overall, similar to the table results, the CTR curve of the GRU algorithm is closer to that of the Planner Oracle compared to the DNN algorithm.

\subsubsection{Conclusions}

From this experiment, we can draw the following conclusions: 1) with explicit multistep planning, the recommender strategy can sacrifice some short-term clicks to get more user engagement in the long run; 2) such recommendation strategy with planning also has a greater impact on user preferences; 3) the recommendation model that utilizing the user's recent behavior is more similar to the multistep planning strategy in both dimensions of CTR and Preference Shift.

\subsection{Data-driven Sequential Recommendation Experiment}

\subsubsection{Experiment Settings}

For experiments of sequential recommendations, we use the user behavior model and accompanying dataset in the recently proposed simulated environment RL4RS~\cite{wang2021rl4rs} benchmark. The dataset in RL4RS is logged from the real user feedback in the most popular games released by \textit{NetEase Games}. The user behavior model is trained on the dataset using DIEN~\cite{zhou2019deep}, which we use as the action model.

Similar to the data-driven slate recommendation experiment, we need to check whether the user behavior model in RL4RS meets \our's requirements. Before the interaction starts, each user in RL4RS is initialized with a certain length of historical clicked items, which is denoted as $C^{0}_i$. At interaction round $r$, the input to the DIEN-based behavior model includes the historical clicked items $C^{r-1}_i$ until round $t-1$, the static user feature $u_i$, the recommended item list $D^r_i=\{d^r_{i1}, \dots, d^r_{im}\}$, and the model outputs the click probability for each document. After each round, the clicked items are appended to the $C^{r-1}_i$ to form $C^{r}_i$. If we view the historical clicked items as part of the user feature, the RL4RS behavior model can fit in the required form:
\begin{align*}
    p(c^r_{ij}=1)&=\text{DIEN}(u_i,  C^{r}_i, d^r_{ij}, D^r_i) \\
    &=\text{DIEN}(u'^{r}_i, d^r_{ij}, D^r_i)~,
\end{align*}
where $u'^r_i=(u_i, C^r_i)$.

We include two oracle strategies for training data collection. Since enumerating all permutations of possible recommendation items becomes infeasible for the large combination space in RL4RS, we use an online RL algorithm with the objective of maximizing the user clicks as the strategy that purely optimizes CTR regardless of FCTR.

\begin{itemize}
    \item \textbf{Unbiased Oracle} is the strategy that always ranks items according to users' initial preference scores.
    \item \textbf{PPO Oracle} is a recommendation policy trained with a reinforcement learning algorithm PPO~\cite{schulman2017proximal} by online interactions with the user model. At each step, the policy picks one item to the list. The reward is sparse and given as the number of clicks on the recommended list at the end of the list.
\end{itemize}

\subsubsection{Evaluated Methods}

We want to investigate whether modeling recent user behavior can lead to a higher Preference Shift. Therefore, we evaluate two sequential recommendation algorithms. For each of them, we implement two versions, Static and Dynamic.

\begin{itemize}
    \item \textbf{GRU Static} uses GRU to encode the users' initial behavior history, and the clicked items during interactions are not appended to the history. 
    \item \textbf{GRU Dynamic} uses GRU to encode the user behavior history, which is actively adapted according to the recently clicked items.
    \item \textbf{DIN~\cite{zhou2018deep} Static} utilizes an attention structure to extract information from user behavior history. Similar to GRU Static, the recent click history is not fed into the model.
    \item \textbf{DIN Dynamic} uses the same network structure and training objective as DIN Static, while the recently clicked items are included in the user behavior history. 
    \item \textbf{BERT4Rec~\cite{sun2019bert4rec} Static} adopts the BERT (Bidirectional Encoder Representations from Transformers) structure to model the user behavior history. To learn the bidirectional representations, models are trained to predict the clicks on randomly masked items in the history using their left and right contexts. The recently clicked items are not appended to the history.
    \item \textbf{BERT4Rec Dynamic} uses the same network structure and training objective as BERT4Rec Static, except the recently clicked items are dynamically updated to the history.
\end{itemize}

\subsubsection{Experiment Results}

\tb{tb:rl4rs-res} presents the benchmarking results when the training data is collected using PPO Oracle. As we look at the offline metrics, three dynamic algorithms have better performance compared to their static counterparts. In online evaluations, results of those dynamic methods that encode the users' recent click history show a similar pattern as reranking methods in slate recommendations. The CTR of DIN Dynamic, GRU Dynamic, and BERT4Rec Dynamic is higher than DIN Static, GRU Static, and BERT4Rec Static, respectively. On the other hand, the FCTR of dynamic methods is relatively lower. Also, regarding the long-term impact on user preference after multiple interaction rounds, we witness that the Preference Shift of dynamic algorithms is also higher. 

\begin{center}
\begin{table}[h]
\begin{tabular}{c|cc|ccc}
    \hline
    \hline
    \multirow{2}*{Method}& \multicolumn{2}{c|}{Offline} & \multicolumn{3}{c}{Online} \\
    ~ & NDCG & AUC & CTR & FCTR & PS\\
    \hline
    GRU Static & 0.6685 & 0.8327 & 0.4643 & \textbf{0.6967} & 0.7575 \\
    GRU Dynamic & \textbf{0.6785} & 0.9297 & 0.4855 & 0.6373 & 0.7655 \\
    DIN Static & 0.6625 & 0.8314 & 0.4534 & 0.6917 & 0.7566 \\
    DIN Dynamic & 0.6673 & 0.9296 & 0.4811 & 0.6738 & 0.7640 \\
    BERT4Rec Static & 0.6649 & 0.8336 & 0.4574 & 0.6947 & \textbf{0.7558} \\
    BERT4Rec Dynamic & 0.6718 & \textbf{0.9330} & \textbf{0.4947} & 0.6106 & 0.7687 \\
    \hline
    Unbiased Oracle & 0.6347 & - & 0.4375 & 0.9998 & 0.6495 \\
    PPO Oracle & 0.6079 & - & 0.5515 & 0.5678 & 0.7736 \\
    \hline
    \hline
\end{tabular}
\vspace{5pt}
\caption{The evaluation results of different algorithms (strategies) on the data-driven sequential recommendation scenario when the mixed ratio is 1.0.}
\label{tb:rl4rs-res}
\end{table}
\end{center}

\begin{figure}[htbp]
    \centering
    \subfigure[]{
        \includegraphics[scale=0.33]{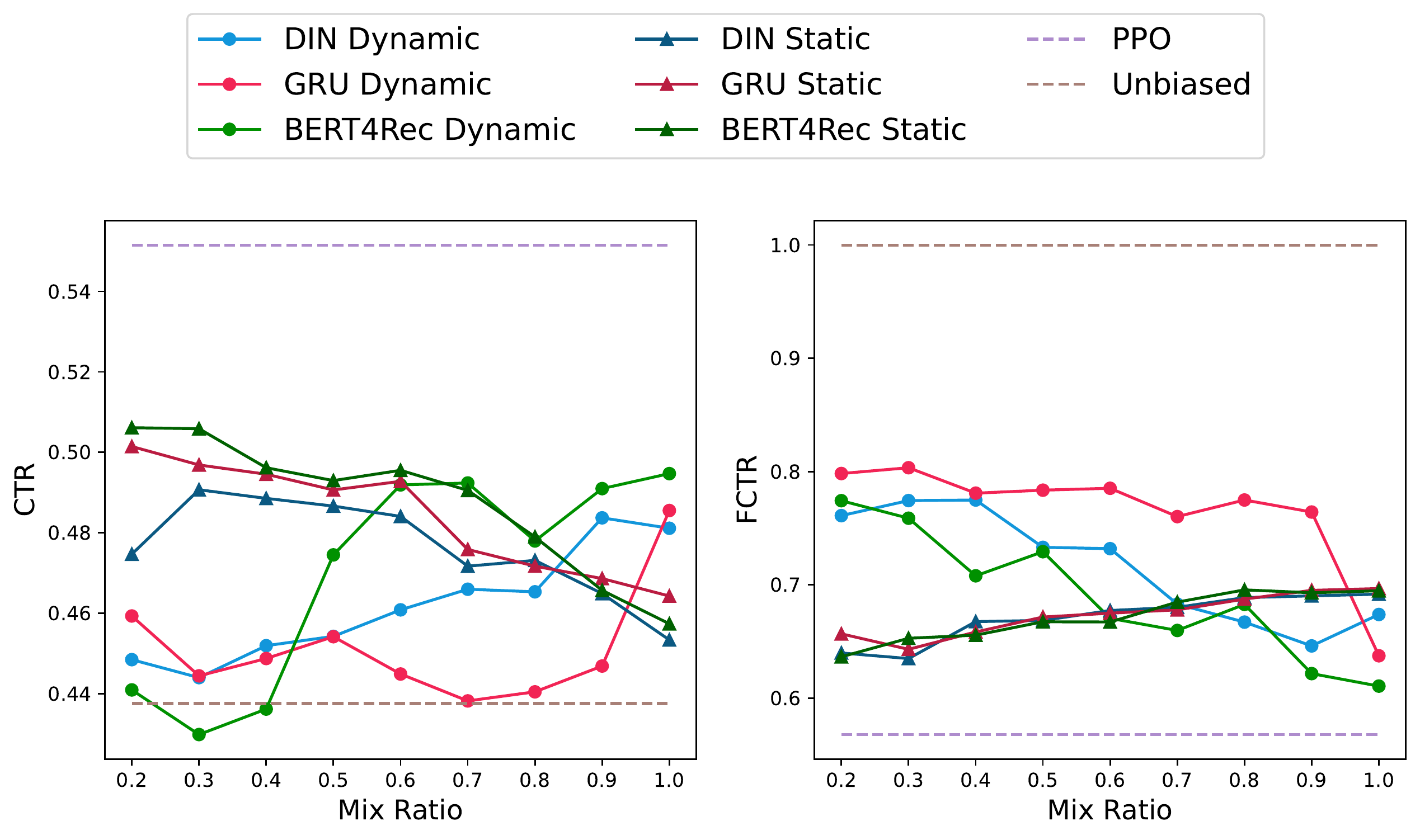}
        \label{fig:seq-mix-line}
    }
    \subfigure[]{
	\includegraphics[scale=0.32]{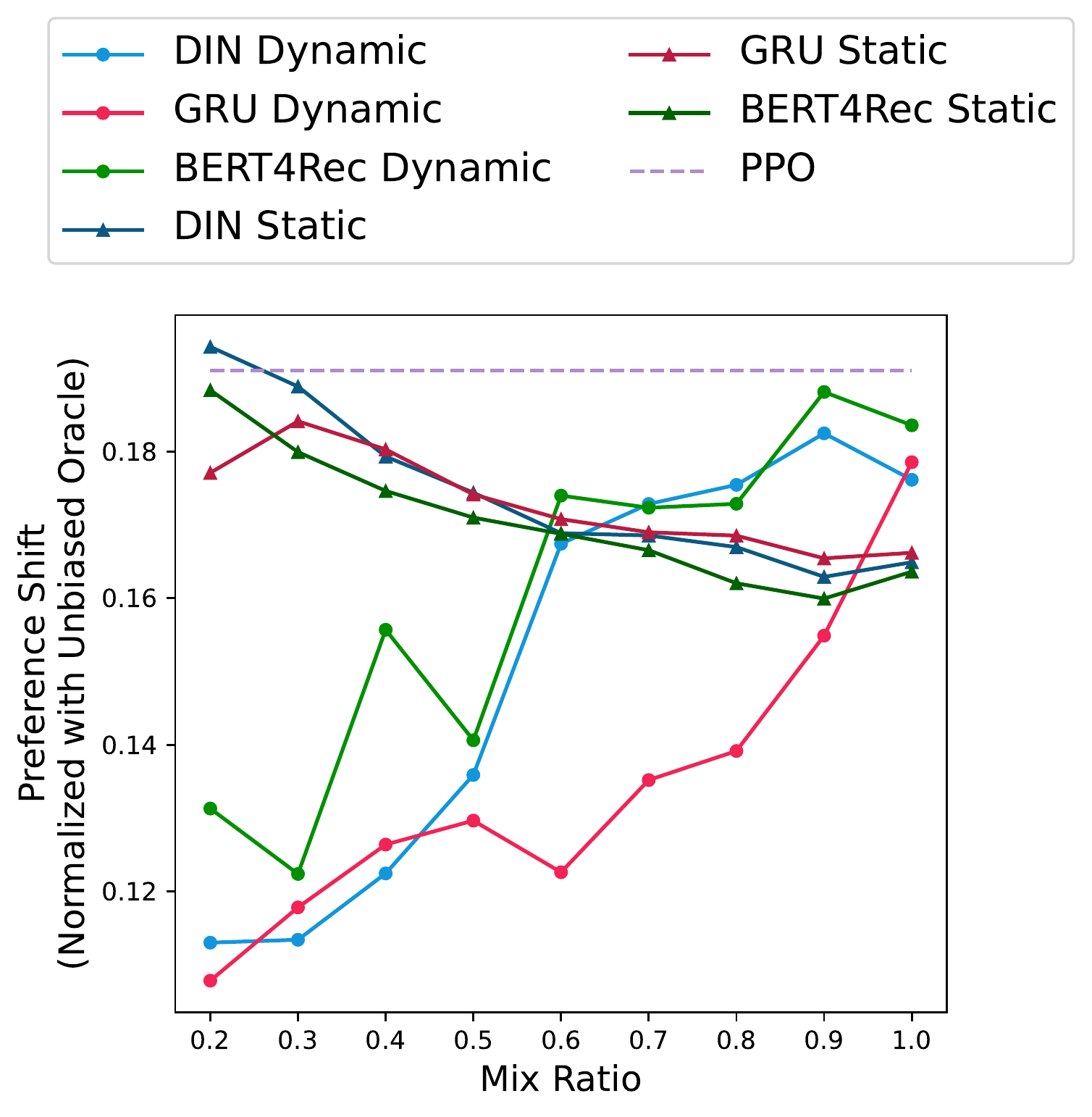}
	\label{fig:seq-preference-shift}
    }
    \caption{Online metrics of difference algorithms (strategies) when trained with data of different mix ratios. Figure (a) shows the changing curves of CTR and FCTR with different mix ratios. Figure (b) shows the changing curves of preference shifts with different mix ratios. For better presentation, the values in Figure (b) are the results of subtracting the values of Unbiased Oracle.}
    \label{fig:rl4rs-mix-ratio}
\end{figure}

In \fig{fig:rl4rs-mix-ratio}, we demonstrate the results when we change the mix ratio of training data from 0.2 to 1.0. When the mix ratio of training data is small, the CTR of GRU Static, DIN Static, and BERT4Rec Static, which have higher CTR and lower FCTR, are more manipulative than dynamic ones. Also, the Preference Shift induced by static algorithms is larger than that induced by dynamic algorithms. This might be due to the property of Unbiased Oracle. For the training data collected by Unbiased Oracle, the recommended slates are not determined by the users' recent history. Also, the user preferences across multiple rounds are also less affected since, from \tb{tb:rl4rs-res}, the Unbiased Oracle has a smaller value of Preference Shift. In such cases, algorithms that take the recent user click history into account could instead overfit the irrelevant information, degrading the online performance. When the mix ratio is close to 1.0, where most of the data is collected by PPO Oracle, the dynamic algorithms have higher CTR but also end up with lower FCTR and higher Preference Shift. 


\subsubsection{Conclusions}
From this experiment, we can draw the following conclusions: 1) real users can be manipulated in sequential recommendations, and such manipulative recommendations cause more significant changes in user preferences; 2) the FCTR and Preference Shift of dynamic algorithms change more dramatically than static algorithms as the mix ratio changes; 3) when the training data is generated exclusively by a manipulative strategy, the dynamic algorithms are more similar to the manipulative strategy in all metrics. 

\subsection{Comprehensive Analysis}
In the final part of the experiment section, we provide a comprehensive analysis of the experimental results and give our general understanding of the manipulations of recommender systems on user preferences.

\subsubsection{Good performance on traditional metrics can come at the cost of manipulating user preferences} 
One of the main reasons we propose this framework is that traditional metrics for evaluating a recommender system, whether online or offline, are agnostic to the system's manipulation of the user's preference. In all four benchmark experiments, we can find some algorithms that are able to achieve higher CTR and perform better on offline metrics like AUC and NDCG but also make users click on a smaller proportion of favorite documents or even actively influence user preferences. For example, in the data-driven slate recommendation scenario, the NDCG and CTR of DeepFM are 0.5835 and 0.2031, respectively. Both are much higher than those of DeepFM, which are 0.5685 and 0.2010. However, when evaluated with manipulation-aware metrics, SetRank is considered to be more manipulative on users' initial preferences, which may negatively affect the reputation of the recommendation platform in the longer term. 

\subsubsection{Data greatly shapes the manipulation degrees of algorithms trained on it} 
In most experiments, we demonstrate the evaluated results when algorithms are trained with different mix ratios of data. The results indicate that the initial strategy or algorithm used to collect the training data significantly impacts on the manipulation behavior of the models trained afterward. For instance, in the data-driven slate commendation experiment, SetRank is always more manipulative than DeepFM, when trained using data with small mix ratios. However, from \fig{fig:slate-mix-line}, we can see that DeepFM trained on the mix ratio of 0.1 is instead more manipulative than SetRank trained on the mix ratio of 0.2. In both synthetic and data-driven experiments, if a highly manipulative algorithm collects more training data, the same machine learning model trained on it tends to make more manipulations on users' preferences. 

\subsubsection{More powerful modeling abilities lead to greater sensitivity to changes in the degree of manipulation within the training data}

We compare recommendation algorithms with different modeling abilities in both slate and sequential recommendation scenarios. Note that the modeling ability here does not refer to the neural model complexity, but the complexity of user and document information it takes into account. In slate recommendations, reranking models explicitly model the interactions of documents within a recommended slate, which are considered to have higher modeling ability compared to point-wise models. Similarly, in sequential recommendations, models like DIN that capture users' recent behaviors have higher modeling ability than static models. From experiments on both data-driven experiments, although the degree of manipulations of all algorithms is affected by the training data, algorithms with stronger modeling abilities undergo greater impact. 

\section{Related Work}
\label{sec:related}

\subsection{The Influence of Recommender Systems on Users and Related Concepts}
Some previous works have shed light on how recommender systems might affect users' internal states or preferences, which is what our research focuses on. 
As is mentioned in ~\cite{carroll2022estimating}, some work is based on empirical study~\cite{chen2019product}; some analyze static datasets of interactions~\cite{nguyen2014exploring, li2014modeling, ribeiro2020auditing}; some are simulation-based, modeling users either using handcrafted, explicit rules~\cite{ie2019recsim, jiang2019degenerate, mansoury2020feedback, ghanem2022balancing, yao2021measuring, chaney2018algorithmic} or learning user dynamics implicitly~\cite{carroll2022estimating}.
Our work is similar to the latter one. With two synthetic and data-driven scenarios, we emphasize the potential impact of RS algorithms on users.

A closely related concept is fairness in recommender systems. While improved online metrics, such as click-through rate, are the most critical metrics for recommender systems, fairness is proposed to ensure recommender systems allocate resources equitably and improve the personal experience and social good. 
In the survey of the fairness of recommender systems~\cite{wang2022survey}, it is noted that both users and items may suffer from unfairness.
For users, considerable differences are found in several recommendation measurements, such as accuracy, diversity, and novelty for users of different ages and genders~\cite{ekstrand2018all, wang2021user}; and for items, less popular items may get less exposure opportunity~\cite{beutel2019fairness, zhu2020measuring}, which is similar to popularity bias~\cite{zhu2021popularity}.
However, the causal relationship is the opposite between our work and the fairness problem.
Our work focuses on how recommender systems influence or manipulate users, providing several metrics to measure the manipulation degree of a particular algorithm under a specific recommendation scenario. In contrast, fairness concerns how different users or items suffer from a specific recommendation system.
The former does not care about the difference between users, and in our work, we take the average score of users instead, while the latter concerns the individuals.

\subsection{User Behavior Modeling}
The modeling of user behavior is an enduring topic in recommender system research. Early attempts were made to build the user click model under the probabilistic graphical model (PGM)~\cite{jordan2003introduction} framework. These models rely heavily on different assumptions to specify how documents in the same slate jointly influence user behavior.
The cascade model (CM)~\cite{craswell2008experimental} adopts the assumption that users scan each document from top to bottom of the slate until the first click. Such an assumption constrains each query session to have exactly one click, and several follow-up works are proposed to overcome the limitation~\cite{srikant2010user, guo2009click}.
Since these simple click models are constrained by strong assumptions, thus are not suitable for our use case of investigating human's bounded rationality in recommendations. 

Neural network-based methods are proposed for more flexible and realistic user behavior modeling. CTR estimation models like DeepFM use neural networks to model the feature interaction of a single item and predict the user click probability of each item independently.
Neural networks are also applied in building click models that take the influences from previous items within the same slate into account.
The neural click model (NCM)~\cite{borisov2016neural} first applies neural networks in building click models.
The click sequence model (CSM)~\cite{borisov2018click} employs an encoder-decoder architecture, in which the encoder computes the contextual embedding of the documents and sends it to the decoder to predict the clicked documents.
Several methods further incorporate the interactions across different slates, i.e., the user behavior history, for better user modeling. DIN~\cite{zhou2018deep} and DIEN~\cite{zhou2019deep} use the attention mechanism to assign different weights to historical behaviors according to their relevance to the target document. Hierarchical Periodic Memory Network (HPMN)~\cite{ren2019lifelong} uses a lifelong personalized memory for storing the user interest representations, which is updated incrementally. User Behavior Retrieval (UBR4CTR)~\cite{qin2020user} uses retrieval techniques to obtain the most relevant behaviors from the entire user behavior sequence.
Another series of works adopt adversarial imitation learning methods to learn user models, and the learned models are then used as environments to train reinforcement-learning-based recommendation algorithms~\cite{shi2019virtual, shang2019environment}. 
Such neural network-based methods can automatically extract the real and complex dependencies within and across different slates from user behavior, including the possible irrational dependencies. 
Since we focus on studying the manipulations based on those existing advanced user models, the details of how the user models are obtained are beyond the scope of our discussion.

\subsection{Simulation-based Research on Recommender Systems}
Measuring algorithms purely on metrics from offline datasets is flawed~\cite{tian2020estimating}, but not all algorithms can be validated online on real recommendation platforms in the first place.
In recent years, some simulation-based platforms have been aiming to provide environments where recommendation algorithms can be evaluated or benchmarked in an online manner, and some work~\cite{bernardi2021simulations} has compared these platforms briefly.
RecoGym~\cite{rohde2018recogym} is a simulation platform that provides a unified recommendation framework for combining classical recommendation algorithms with reinforcement learning methods, allowing both offline and online experiments. Still, it can be far from the real-world dataset by just using synthetic data.
PyRecGym~\cite{shi2019pyrecgym}  extends RecoGym by supporting multiple input data types and user-feedback functions, but its architecture makes it difficult to create reusable simulation environments.
RecSim~\cite{ie2019recsim}, like RecoGym, focuses on reinforcement learning methods for recommender systems and is suitable for simulating continuous interactive recommendation problems.
It allows creating environments that can reflect certain aspects of user behavior but it is still hard to simulate a complex system.
Virtual-taobao~\cite{shi2019virtual} can reflect properties of the real environment by generating virtual customers and interactions through GANSD and MAIL, respectively.
To address the gap between simulation and real-world applications, RecSim NG~\cite{mladenov2020demonstrating}, a scalable, modular, differentiable simulator, is proposed based on RecSim, offering a powerful, general probabilistic programming language for agent-behavior specification.
However, the evaluation criteria adopted in existing works are aligned with traditional online metrics, and no simulation framework explicitly measures the degree to which recommendation algorithms manipulate user preferences.
Our work lies in the general area of simulation-based study of recommender systems using real-world datasets, but investigates a specific problem, manipulations of users' preferences from recommender systems, which challenges the effectiveness of traditional online metrics. 

\section{Conclusion and Future Work}
\label{sec:conclusion}
In this work, we put forward and investigate the problem of manipulations of users' preferences in today's recommender systems. Based on the bounded rationality of human behavior, we challenge the sources of advances in the online performance of current recommendation algorithms. We posit that the improvement in online performance may come from manipulating user preferences, and propose a general and configurable benchmark framework named \our to measure the manipulations. We make the simulated user behavior model to provide individual scores for documents in the entire document set one by one, which serves as the core of our manipulation-aware evaluations. 


\paragraph{Key Findings}
We use our proposed framework to evaluate various recommendation algorithms in both slate and sequential recommendations. We first find that in slate recommendations, more advanced reranking algorithms over-exploit the influence of the interrelationship of documents within a slate on human decisions, which causes users to choose more documents they do not favor, despite the uplift in traditional online metrics. In sequential recommendations, those algorithms that dynamically model the user's recent behavior will significantly impact on user preference, resulting in users choosing less favored documents. Moreover, in all scenarios, we found that the degree of manipulation in the training data had a significant effect on the degree of manipulation of the final recommendation model. 


\paragraph{Future Work}
Our proposed framework requires access to the raw output of ranking models and the complete document set, which is not available for each individual user. In future work, we will study the approximated estimation of the manipulations from the user side, which can be used as personal supervision of recommender systems. 
Moreover, we are interested in proposing practical solutions to reduce manipulations in recommendation systems. We also hope that our proposed benchmark will make the academic community and the industry aware of the limitations of existing online metrics and be concerned about the negative impact of manipulations on user preferences from the recommender systems. 

\begin{acks}
The Shanghai Jiao Tong University team is supported by National Key R\&D Program of China (2022ZD0114804), Shanghai Municipal Science and Technology Major Project (2021SHZDZX0102) and National Natural Science Foundation of China (62177033). The Nanjing University team is supported by National Natural Science Foundation of China (61921006).
\end{acks}

\bibliographystyle{ACM-Reference-Format}
\bibliography{ref}


\begin{thebibliography}{102}


\ifx \showCODEN    \undefined \def \showCODEN     #1{\unskip}     \fi
\ifx \showDOI      \undefined \def \showDOI       #1{#1}\fi
\ifx \showISBNx    \undefined \def \showISBNx     #1{\unskip}     \fi
\ifx \showISBNxiii \undefined \def \showISBNxiii  #1{\unskip}     \fi
\ifx \showISSN     \undefined \def \showISSN      #1{\unskip}     \fi
\ifx \showLCCN     \undefined \def \showLCCN      #1{\unskip}     \fi
\ifx \shownote     \undefined \def \shownote      #1{#1}          \fi
\ifx \showarticletitle \undefined \def \showarticletitle #1{#1}   \fi
\ifx \showURL      \undefined \def \showURL       {\relax}        \fi
\providecommand\bibfield[2]{#2}
\providecommand\bibinfo[2]{#2}
\providecommand\natexlab[1]{#1}
\providecommand\showeprint[2][]{arXiv:#2}

\bibitem[Adomavicius et~al\mbox{.}(2018)]%
        {adomavicius2018effects}
\bibfield{author}{\bibinfo{person}{Gediminas Adomavicius}, \bibinfo{person}{Jesse~C Bockstedt}, \bibinfo{person}{Shawn~P Curley}, {and} \bibinfo{person}{Jingjing Zhang}.} \bibinfo{year}{2018}\natexlab{}.
\newblock \showarticletitle{Effects of online recommendations on consumers’ willingness to pay}.
\newblock \bibinfo{journal}{\emph{Information Systems Research}} \bibinfo{volume}{29}, \bibinfo{number}{1} (\bibinfo{year}{2018}), \bibinfo{pages}{84--102}.
\newblock


\bibitem[Ai et~al\mbox{.}(2018)]%
        {ai2018learning}
\bibfield{author}{\bibinfo{person}{Qingyao Ai}, \bibinfo{person}{Keping Bi}, \bibinfo{person}{Jiafeng Guo}, {and} \bibinfo{person}{W~Bruce Croft}.} \bibinfo{year}{2018}\natexlab{}.
\newblock \showarticletitle{Learning a deep listwise context model for ranking refinement}. In \bibinfo{booktitle}{\emph{The 41st international ACM SIGIR conference on research \& development in information retrieval}}. \bibinfo{pages}{135--144}.
\newblock


\bibitem[Bergman et~al\mbox{.}(2010)]%
        {bergman2010anchoring}
\bibfield{author}{\bibinfo{person}{Oscar Bergman}, \bibinfo{person}{Tore Ellingsen}, \bibinfo{person}{Magnus Johannesson}, {and} \bibinfo{person}{Cicek Svensson}.} \bibinfo{year}{2010}\natexlab{}.
\newblock \showarticletitle{Anchoring and cognitive ability}.
\newblock \bibinfo{journal}{\emph{Economics Letters}} \bibinfo{volume}{107}, \bibinfo{number}{1} (\bibinfo{year}{2010}), \bibinfo{pages}{66--68}.
\newblock


\bibitem[Bernardi et~al\mbox{.}(2021)]%
        {bernardi2021simulations}
\bibfield{author}{\bibinfo{person}{Lucas Bernardi}, \bibinfo{person}{Sakshi Batra}, {and} \bibinfo{person}{Cintia~Alicia Bruscantini}.} \bibinfo{year}{2021}\natexlab{}.
\newblock \showarticletitle{Simulations in Recommender Systems: An industry perspective}.
\newblock \bibinfo{journal}{\emph{arXiv preprint arXiv:2109.06723}} (\bibinfo{year}{2021}).
\newblock


\bibitem[Beutel et~al\mbox{.}(2019)]%
        {beutel2019fairness}
\bibfield{author}{\bibinfo{person}{Alex Beutel}, \bibinfo{person}{Jilin Chen}, \bibinfo{person}{Tulsee Doshi}, \bibinfo{person}{Hai Qian}, \bibinfo{person}{Li Wei}, \bibinfo{person}{Yi Wu}, \bibinfo{person}{Lukasz Heldt}, \bibinfo{person}{Zhe Zhao}, \bibinfo{person}{Lichan Hong}, \bibinfo{person}{Ed~H Chi}, {et~al\mbox{.}}} \bibinfo{year}{2019}\natexlab{}.
\newblock \showarticletitle{Fairness in recommendation ranking through pairwise comparisons}. In \bibinfo{booktitle}{\emph{Proceedings of the 25th ACM SIGKDD International Conference on Knowledge Discovery \& Data Mining}}. \bibinfo{pages}{2212--2220}.
\newblock


\bibitem[Beutel et~al\mbox{.}(2018)]%
        {beutel2018latent}
\bibfield{author}{\bibinfo{person}{Alex Beutel}, \bibinfo{person}{Paul Covington}, \bibinfo{person}{Sagar Jain}, \bibinfo{person}{Can Xu}, \bibinfo{person}{Jia Li}, \bibinfo{person}{Vince Gatto}, {and} \bibinfo{person}{Ed~H Chi}.} \bibinfo{year}{2018}\natexlab{}.
\newblock \showarticletitle{Latent cross: Making use of context in recurrent recommender systems}. In \bibinfo{booktitle}{\emph{Proceedings of the Eleventh ACM International Conference on Web Search and Data Mining}}. \bibinfo{pages}{46--54}.
\newblock


\bibitem[Borisov et~al\mbox{.}(2016)]%
        {borisov2016neural}
\bibfield{author}{\bibinfo{person}{Alexey Borisov}, \bibinfo{person}{Ilya Markov}, \bibinfo{person}{Maarten De~Rijke}, {and} \bibinfo{person}{Pavel Serdyukov}.} \bibinfo{year}{2016}\natexlab{}.
\newblock \showarticletitle{A neural click model for web search}. In \bibinfo{booktitle}{\emph{Proceedings of the 25th International Conference on World Wide Web}}. \bibinfo{pages}{531--541}.
\newblock


\bibitem[Borisov et~al\mbox{.}(2018)]%
        {borisov2018click}
\bibfield{author}{\bibinfo{person}{Alexey Borisov}, \bibinfo{person}{Martijn Wardenaar}, \bibinfo{person}{Ilya Markov}, {and} \bibinfo{person}{Maarten de Rijke}.} \bibinfo{year}{2018}\natexlab{}.
\newblock \showarticletitle{A click sequence model for web search}. In \bibinfo{booktitle}{\emph{The 41st International ACM SIGIR Conference on Research \& Development in Information Retrieval}}. \bibinfo{pages}{45--54}.
\newblock


\bibitem[Burges et~al\mbox{.}(2005)]%
        {burges2005learning}
\bibfield{author}{\bibinfo{person}{Chris Burges}, \bibinfo{person}{Tal Shaked}, \bibinfo{person}{Erin Renshaw}, \bibinfo{person}{Ari Lazier}, \bibinfo{person}{Matt Deeds}, \bibinfo{person}{Nicole Hamilton}, {and} \bibinfo{person}{Greg Hullender}.} \bibinfo{year}{2005}\natexlab{}.
\newblock \showarticletitle{Learning to rank using gradient descent}. In \bibinfo{booktitle}{\emph{Proceedings of the 22nd international conference on Machine learning}}. \bibinfo{pages}{89--96}.
\newblock


\bibitem[Burges(2010)]%
        {burges2010ranknet}
\bibfield{author}{\bibinfo{person}{Christopher~JC Burges}.} \bibinfo{year}{2010}\natexlab{}.
\newblock \showarticletitle{From ranknet to lambdarank to lambdamart: An overview}.
\newblock \bibinfo{journal}{\emph{Learning}} \bibinfo{volume}{11}, \bibinfo{number}{23-581} (\bibinfo{year}{2010}), \bibinfo{pages}{81}.
\newblock


\bibitem[Camerer(1999)]%
        {camerer1999behavioral}
\bibfield{author}{\bibinfo{person}{Colin Camerer}.} \bibinfo{year}{1999}\natexlab{}.
\newblock \showarticletitle{Behavioral economics: Reunifying psychology and economics}.
\newblock \bibinfo{journal}{\emph{Proceedings of the National Academy of Sciences}} \bibinfo{volume}{96}, \bibinfo{number}{19} (\bibinfo{year}{1999}), \bibinfo{pages}{10575--10577}.
\newblock


\bibitem[Carroll et~al\mbox{.}(2022)]%
        {carroll2022estimating}
\bibfield{author}{\bibinfo{person}{Micah Carroll}, \bibinfo{person}{Dylan Hadfield-Menell}, \bibinfo{person}{Stuart Russell}, {and} \bibinfo{person}{Anca Dragan}.} \bibinfo{year}{2022}\natexlab{}.
\newblock \showarticletitle{Estimating and Penalizing Induced Preference Shifts in Recommender Systems}.
\newblock \bibinfo{journal}{\emph{arXiv preprint arXiv:2204.11966}} (\bibinfo{year}{2022}).
\newblock


\bibitem[Chaney et~al\mbox{.}(2018)]%
        {chaney2018algorithmic}
\bibfield{author}{\bibinfo{person}{Allison~JB Chaney}, \bibinfo{person}{Brandon~M Stewart}, {and} \bibinfo{person}{Barbara~E Engelhardt}.} \bibinfo{year}{2018}\natexlab{}.
\newblock \showarticletitle{How algorithmic confounding in recommendation systems increases homogeneity and decreases utility}. In \bibinfo{booktitle}{\emph{Proceedings of the 12th ACM conference on recommender systems}}. \bibinfo{pages}{224--232}.
\newblock


\bibitem[Charness and Dave(2017)]%
        {charness2017confirmation}
\bibfield{author}{\bibinfo{person}{Gary Charness} {and} \bibinfo{person}{Chetan Dave}.} \bibinfo{year}{2017}\natexlab{}.
\newblock \showarticletitle{Confirmation bias with motivated beliefs}.
\newblock \bibinfo{journal}{\emph{Games and Economic Behavior}}  \bibinfo{volume}{104} (\bibinfo{year}{2017}), \bibinfo{pages}{1--23}.
\newblock


\bibitem[Chen et~al\mbox{.}(2019b)]%
        {chen2019tiangong}
\bibfield{author}{\bibinfo{person}{Jia Chen}, \bibinfo{person}{Jiaxin Mao}, \bibinfo{person}{Yiqun Liu}, \bibinfo{person}{Min Zhang}, {and} \bibinfo{person}{Shaoping Ma}.} \bibinfo{year}{2019}\natexlab{b}.
\newblock \showarticletitle{TianGong-ST: A new dataset with large-scale refined real-world web search sessions}. In \bibinfo{booktitle}{\emph{Proceedings of the 28th ACM International Conference on Information and Knowledge Management}}. \bibinfo{pages}{2485--2488}.
\newblock


\bibitem[Chen et~al\mbox{.}(2020)]%
        {chen2020context}
\bibfield{author}{\bibinfo{person}{Jia Chen}, \bibinfo{person}{Jiaxin Mao}, \bibinfo{person}{Yiqun Liu}, \bibinfo{person}{Min Zhang}, {and} \bibinfo{person}{Shaoping Ma}.} \bibinfo{year}{2020}\natexlab{}.
\newblock \showarticletitle{A context-aware click model for web search}. In \bibinfo{booktitle}{\emph{Proceedings of the 13th International Conference on Web Search and Data Mining}}. \bibinfo{pages}{88--96}.
\newblock


\bibitem[Chen et~al\mbox{.}(2019c)]%
        {chen2019behavior}
\bibfield{author}{\bibinfo{person}{Qiwei Chen}, \bibinfo{person}{Huan Zhao}, \bibinfo{person}{Wei Li}, \bibinfo{person}{Pipei Huang}, {and} \bibinfo{person}{Wenwu Ou}.} \bibinfo{year}{2019}\natexlab{c}.
\newblock \showarticletitle{Behavior sequence transformer for e-commerce recommendation in alibaba}. In \bibinfo{booktitle}{\emph{Proceedings of the 1st International Workshop on Deep Learning Practice for High-Dimensional Sparse Data}}. \bibinfo{pages}{1--4}.
\newblock


\bibitem[Chen and Canny(2011)]%
        {chen2011recommending}
\bibfield{author}{\bibinfo{person}{Ye Chen} {and} \bibinfo{person}{John~F Canny}.} \bibinfo{year}{2011}\natexlab{}.
\newblock \showarticletitle{Recommending ephemeral items at web scale}. In \bibinfo{booktitle}{\emph{Proceedings of the 34th international ACM SIGIR conference on Research and development in Information Retrieval}}. \bibinfo{pages}{1013--1022}.
\newblock


\bibitem[Chen et~al\mbox{.}(2019a)]%
        {chen2019product}
\bibfield{author}{\bibinfo{person}{Yanhong Chen}, \bibinfo{person}{Yaobin Lu}, \bibinfo{person}{Bin Wang}, {and} \bibinfo{person}{Zhao Pan}.} \bibinfo{year}{2019}\natexlab{a}.
\newblock \showarticletitle{How do product recommendations affect impulse buying? An empirical study on WeChat social commerce}.
\newblock \bibinfo{journal}{\emph{Information \& Management}} \bibinfo{volume}{56}, \bibinfo{number}{2} (\bibinfo{year}{2019}), \bibinfo{pages}{236--248}.
\newblock


\bibitem[Cheng et~al\mbox{.}(2016)]%
        {cheng2016wide}
\bibfield{author}{\bibinfo{person}{Heng-Tze Cheng}, \bibinfo{person}{Levent Koc}, \bibinfo{person}{Jeremiah Harmsen}, \bibinfo{person}{Tal Shaked}, \bibinfo{person}{Tushar Chandra}, \bibinfo{person}{Hrishi Aradhye}, \bibinfo{person}{Glen Anderson}, \bibinfo{person}{Greg Corrado}, \bibinfo{person}{Wei Chai}, \bibinfo{person}{Mustafa Ispir}, {et~al\mbox{.}}} \bibinfo{year}{2016}\natexlab{}.
\newblock \showarticletitle{Wide \& deep learning for recommender systems}. In \bibinfo{booktitle}{\emph{Proceedings of the 1st workshop on deep learning for recommender systems}}. \bibinfo{pages}{7--10}.
\newblock


\bibitem[Cho et~al\mbox{.}(2014)]%
        {cho2014properties}
\bibfield{author}{\bibinfo{person}{Kyunghyun Cho}, \bibinfo{person}{Bart Van~Merri{\"e}nboer}, \bibinfo{person}{Dzmitry Bahdanau}, {and} \bibinfo{person}{Yoshua Bengio}.} \bibinfo{year}{2014}\natexlab{}.
\newblock \showarticletitle{On the properties of neural machine translation: Encoder-decoder approaches}.
\newblock \bibinfo{journal}{\emph{arXiv preprint arXiv:1409.1259}} (\bibinfo{year}{2014}).
\newblock


\bibitem[Chorus et~al\mbox{.}(2014)]%
        {chorus2014random}
\bibfield{author}{\bibinfo{person}{Caspar Chorus}, \bibinfo{person}{Sander van Cranenburgh}, {and} \bibinfo{person}{Thijs Dekker}.} \bibinfo{year}{2014}\natexlab{}.
\newblock \showarticletitle{Random regret minimization for consumer choice modeling: Assessment of empirical evidence}.
\newblock \bibinfo{journal}{\emph{Journal of Business Research}} \bibinfo{volume}{67}, \bibinfo{number}{11} (\bibinfo{year}{2014}), \bibinfo{pages}{2428--2436}.
\newblock


\bibitem[Craswell et~al\mbox{.}(2008)]%
        {craswell2008experimental}
\bibfield{author}{\bibinfo{person}{Nick Craswell}, \bibinfo{person}{Onno Zoeter}, \bibinfo{person}{Michael Taylor}, {and} \bibinfo{person}{Bill Ramsey}.} \bibinfo{year}{2008}\natexlab{}.
\newblock \showarticletitle{An experimental comparison of click position-bias models}. In \bibinfo{booktitle}{\emph{Proceedings of the 2008 international conference on web search and data mining}}. \bibinfo{pages}{87--94}.
\newblock


\bibitem[Dai et~al\mbox{.}(2021)]%
        {dai2021adversarial}
\bibfield{author}{\bibinfo{person}{Xinyi Dai}, \bibinfo{person}{Jianghao Lin}, \bibinfo{person}{Weinan Zhang}, \bibinfo{person}{Shuai Li}, \bibinfo{person}{Weiwen Liu}, \bibinfo{person}{Ruiming Tang}, \bibinfo{person}{Xiuqiang He}, \bibinfo{person}{Jianye Hao}, \bibinfo{person}{Jun Wang}, {and} \bibinfo{person}{Yong Yu}.} \bibinfo{year}{2021}\natexlab{}.
\newblock \showarticletitle{An Adversarial Imitation Click Model for Information Retrieval}. In \bibinfo{booktitle}{\emph{Proceedings of the Web Conference 2021}}. \bibinfo{pages}{1809--1820}.
\newblock


\bibitem[Das et~al\mbox{.}(2007)]%
        {das2007google}
\bibfield{author}{\bibinfo{person}{Abhinandan~S Das}, \bibinfo{person}{Mayur Datar}, \bibinfo{person}{Ashutosh Garg}, {and} \bibinfo{person}{Shyam Rajaram}.} \bibinfo{year}{2007}\natexlab{}.
\newblock \showarticletitle{Google news personalization: scalable online collaborative filtering}. In \bibinfo{booktitle}{\emph{Proceedings of the 16th international conference on World Wide Web}}. \bibinfo{pages}{271--280}.
\newblock


\bibitem[Ekstrand et~al\mbox{.}(2018)]%
        {ekstrand2018all}
\bibfield{author}{\bibinfo{person}{Michael~D Ekstrand}, \bibinfo{person}{Mucun Tian}, \bibinfo{person}{Ion~Madrazo Azpiazu}, \bibinfo{person}{Jennifer~D Ekstrand}, \bibinfo{person}{Oghenemaro Anuyah}, \bibinfo{person}{David McNeill}, {and} \bibinfo{person}{Maria~Soledad Pera}.} \bibinfo{year}{2018}\natexlab{}.
\newblock \showarticletitle{All the cool kids, how do they fit in?: Popularity and demographic biases in recommender evaluation and effectiveness}. In \bibinfo{booktitle}{\emph{Conference on fairness, accountability and transparency}}. PMLR, \bibinfo{pages}{172--186}.
\newblock


\bibitem[Fan et~al\mbox{.}(2020)]%
        {fan2020interpretability}
\bibfield{author}{\bibinfo{person}{Fenglei Fan}, \bibinfo{person}{Jinjun Xiong}, {and} \bibinfo{person}{Ge Wang}.} \bibinfo{year}{2020}\natexlab{}.
\newblock \showarticletitle{On interpretability of artificial neural networks}.
\newblock \bibinfo{journal}{\emph{arXiv preprint arXiv:2001.02522}} \bibinfo{volume}{2}, \bibinfo{number}{10} (\bibinfo{year}{2020}).
\newblock


\bibitem[Fujimoto et~al\mbox{.}(2019)]%
        {fujimoto2019off}
\bibfield{author}{\bibinfo{person}{Scott Fujimoto}, \bibinfo{person}{David Meger}, {and} \bibinfo{person}{Doina Precup}.} \bibinfo{year}{2019}\natexlab{}.
\newblock \showarticletitle{Off-policy deep reinforcement learning without exploration}. In \bibinfo{booktitle}{\emph{International Conference on Machine Learning}}. PMLR, \bibinfo{pages}{2052--2062}.
\newblock


\bibitem[Furnham and Boo(2011)]%
        {furnham2011literature}
\bibfield{author}{\bibinfo{person}{Adrian Furnham} {and} \bibinfo{person}{Hua~Chu Boo}.} \bibinfo{year}{2011}\natexlab{}.
\newblock \showarticletitle{A literature review of the anchoring effect}.
\newblock \bibinfo{journal}{\emph{The journal of socio-economics}} \bibinfo{volume}{40}, \bibinfo{number}{1} (\bibinfo{year}{2011}), \bibinfo{pages}{35--42}.
\newblock


\bibitem[Garg et~al\mbox{.}(2020)]%
        {garg2020batch}
\bibfield{author}{\bibinfo{person}{Diksha Garg}, \bibinfo{person}{Priyanka Gupta}, \bibinfo{person}{Pankaj Malhotra}, \bibinfo{person}{Lovekesh Vig}, {and} \bibinfo{person}{Gautam Shroff}.} \bibinfo{year}{2020}\natexlab{}.
\newblock \showarticletitle{Batch-Constrained Distributional Reinforcement Learning for Session-based Recommendation}.
\newblock \bibinfo{journal}{\emph{arXiv preprint arXiv:2012.08984}} (\bibinfo{year}{2020}).
\newblock


\bibitem[Ghanem et~al\mbox{.}(2022)]%
        {ghanem2022balancing}
\bibfield{author}{\bibinfo{person}{Nada Ghanem}, \bibinfo{person}{Stephan Leitner}, {and} \bibinfo{person}{Dietmar Jannach}.} \bibinfo{year}{2022}\natexlab{}.
\newblock \showarticletitle{Balancing Consumer and Business Value of Recommender Systems: A Simulation-based Analysis}.
\newblock \bibinfo{journal}{\emph{arXiv preprint arXiv:2203.05952}} (\bibinfo{year}{2022}).
\newblock


\bibitem[Guevara and Fukushi(2016)]%
        {guevara2016modeling}
\bibfield{author}{\bibinfo{person}{C~Angelo Guevara} {and} \bibinfo{person}{Mitsuyoshi Fukushi}.} \bibinfo{year}{2016}\natexlab{}.
\newblock \showarticletitle{Modeling the decoy effect with context-RUM Models: Diagrammatic analysis and empirical evidence from route choice SP and mode choice RP case studies}.
\newblock \bibinfo{journal}{\emph{Transportation Research Part B: Methodological}}  \bibinfo{volume}{93} (\bibinfo{year}{2016}), \bibinfo{pages}{318--337}.
\newblock


\bibitem[Guo et~al\mbox{.}(2009)]%
        {guo2009click}
\bibfield{author}{\bibinfo{person}{Fan Guo}, \bibinfo{person}{Chao Liu}, \bibinfo{person}{Anitha Kannan}, \bibinfo{person}{Tom Minka}, \bibinfo{person}{Michael Taylor}, \bibinfo{person}{Yi-Min Wang}, {and} \bibinfo{person}{Christos Faloutsos}.} \bibinfo{year}{2009}\natexlab{}.
\newblock \showarticletitle{Click chain model in web search}. In \bibinfo{booktitle}{\emph{Proceedings of the 18th international conference on World wide web}}. \bibinfo{pages}{11--20}.
\newblock


\bibitem[Guo et~al\mbox{.}(2017)]%
        {guo2017deepfm}
\bibfield{author}{\bibinfo{person}{Huifeng Guo}, \bibinfo{person}{Ruiming Tang}, \bibinfo{person}{Yunming Ye}, \bibinfo{person}{Zhenguo Li}, {and} \bibinfo{person}{Xiuqiang He}.} \bibinfo{year}{2017}\natexlab{}.
\newblock \showarticletitle{DeepFM: a factorization-machine based neural network for CTR prediction}.
\newblock \bibinfo{journal}{\emph{arXiv preprint arXiv:1703.04247}} (\bibinfo{year}{2017}).
\newblock


\bibitem[He et~al\mbox{.}(2016)]%
        {he2016vista}
\bibfield{author}{\bibinfo{person}{Ruining He}, \bibinfo{person}{Chen Fang}, \bibinfo{person}{Zhaowen Wang}, {and} \bibinfo{person}{Julian McAuley}.} \bibinfo{year}{2016}\natexlab{}.
\newblock \showarticletitle{Vista: A visually, socially, and temporally-aware model for artistic recommendation}. In \bibinfo{booktitle}{\emph{Proceedings of the 10th acm conference on recommender systems}}. \bibinfo{pages}{309--316}.
\newblock


\bibitem[He and McAuley(2016)]%
        {he2016fusing}
\bibfield{author}{\bibinfo{person}{Ruining He} {and} \bibinfo{person}{Julian McAuley}.} \bibinfo{year}{2016}\natexlab{}.
\newblock \showarticletitle{Fusing similarity models with markov chains for sparse sequential recommendation}. In \bibinfo{booktitle}{\emph{2016 IEEE 16th International Conference on Data Mining (ICDM)}}. IEEE, \bibinfo{pages}{191--200}.
\newblock


\bibitem[He et~al\mbox{.}(2018)]%
        {he2018outer}
\bibfield{author}{\bibinfo{person}{Xiangnan He}, \bibinfo{person}{Xiaoyu Du}, \bibinfo{person}{Xiang Wang}, \bibinfo{person}{Feng Tian}, \bibinfo{person}{Jinhui Tang}, {and} \bibinfo{person}{Tat-Seng Chua}.} \bibinfo{year}{2018}\natexlab{}.
\newblock \showarticletitle{Outer product-based neural collaborative filtering}.
\newblock \bibinfo{journal}{\emph{arXiv preprint arXiv:1808.03912}} (\bibinfo{year}{2018}).
\newblock


\bibitem[He et~al\mbox{.}(2017)]%
        {he2017neural}
\bibfield{author}{\bibinfo{person}{Xiangnan He}, \bibinfo{person}{Lizi Liao}, \bibinfo{person}{Hanwang Zhang}, \bibinfo{person}{Liqiang Nie}, \bibinfo{person}{Xia Hu}, {and} \bibinfo{person}{Tat-Seng Chua}.} \bibinfo{year}{2017}\natexlab{}.
\newblock \showarticletitle{Neural collaborative filtering}. In \bibinfo{booktitle}{\emph{Proceedings of the 26th international conference on world wide web}}. \bibinfo{pages}{173--182}.
\newblock


\bibitem[Hidasi and Karatzoglou(2018)]%
        {hidasi2018recurrent}
\bibfield{author}{\bibinfo{person}{Bal{\'a}zs Hidasi} {and} \bibinfo{person}{Alexandros Karatzoglou}.} \bibinfo{year}{2018}\natexlab{}.
\newblock \showarticletitle{Recurrent neural networks with top-k gains for session-based recommendations}. In \bibinfo{booktitle}{\emph{Proceedings of the 27th ACM international conference on information and knowledge management}}. \bibinfo{pages}{843--852}.
\newblock


\bibitem[Hidasi et~al\mbox{.}(2015)]%
        {hidasi2015session}
\bibfield{author}{\bibinfo{person}{Bal{\'a}zs Hidasi}, \bibinfo{person}{Alexandros Karatzoglou}, \bibinfo{person}{Linas Baltrunas}, {and} \bibinfo{person}{Domonkos Tikk}.} \bibinfo{year}{2015}\natexlab{}.
\newblock \showarticletitle{Session-based recommendations with recurrent neural networks}.
\newblock \bibinfo{journal}{\emph{arXiv preprint arXiv:1511.06939}} (\bibinfo{year}{2015}).
\newblock


\bibitem[Huber et~al\mbox{.}(1982)]%
        {huber1982adding}
\bibfield{author}{\bibinfo{person}{Joel Huber}, \bibinfo{person}{John~W Payne}, {and} \bibinfo{person}{Christopher Puto}.} \bibinfo{year}{1982}\natexlab{}.
\newblock \showarticletitle{Adding asymmetrically dominated alternatives: Violations of regularity and the similarity hypothesis}.
\newblock \bibinfo{journal}{\emph{Journal of consumer research}} \bibinfo{volume}{9}, \bibinfo{number}{1} (\bibinfo{year}{1982}), \bibinfo{pages}{90--98}.
\newblock


\bibitem[Ie et~al\mbox{.}(2019a)]%
        {ie2019recsim}
\bibfield{author}{\bibinfo{person}{Eugene Ie}, \bibinfo{person}{Chih-wei Hsu}, \bibinfo{person}{Martin Mladenov}, \bibinfo{person}{Vihan Jain}, \bibinfo{person}{Sanmit Narvekar}, \bibinfo{person}{Jing Wang}, \bibinfo{person}{Rui Wu}, {and} \bibinfo{person}{Craig Boutilier}.} \bibinfo{year}{2019}\natexlab{a}.
\newblock \showarticletitle{Recsim: A configurable simulation platform for recommender systems}.
\newblock \bibinfo{journal}{\emph{arXiv preprint arXiv:1909.04847}} (\bibinfo{year}{2019}).
\newblock


\bibitem[Ie et~al\mbox{.}(2019b)]%
        {ie2019slateq}
\bibfield{author}{\bibinfo{person}{Eugene Ie}, \bibinfo{person}{Vihan Jain}, \bibinfo{person}{Jing Wang}, \bibinfo{person}{Sanmit Narvekar}, \bibinfo{person}{Ritesh Agarwal}, \bibinfo{person}{Rui Wu}, \bibinfo{person}{Heng-Tze Cheng}, \bibinfo{person}{Tushar Chandra}, {and} \bibinfo{person}{Craig Boutilier}.} \bibinfo{year}{2019}\natexlab{b}.
\newblock \showarticletitle{SlateQ: A tractable decomposition for reinforcement learning with recommendation sets}.
\newblock  (\bibinfo{year}{2019}).
\newblock


\bibitem[Jiang et~al\mbox{.}(2019)]%
        {jiang2019degenerate}
\bibfield{author}{\bibinfo{person}{Ray Jiang}, \bibinfo{person}{Silvia Chiappa}, \bibinfo{person}{Tor Lattimore}, \bibinfo{person}{Andr{\'a}s Gy{\"o}rgy}, {and} \bibinfo{person}{Pushmeet Kohli}.} \bibinfo{year}{2019}\natexlab{}.
\newblock \showarticletitle{Degenerate feedback loops in recommender systems}. In \bibinfo{booktitle}{\emph{Proceedings of the 2019 AAAI/ACM Conference on AI, Ethics, and Society}}. \bibinfo{pages}{383--390}.
\newblock


\bibitem[Jing and Smola(2017)]%
        {jing2017neural}
\bibfield{author}{\bibinfo{person}{How Jing} {and} \bibinfo{person}{Alexander~J Smola}.} \bibinfo{year}{2017}\natexlab{}.
\newblock \showarticletitle{Neural survival recommender}. In \bibinfo{booktitle}{\emph{Proceedings of the Tenth ACM International Conference on Web Search and Data Mining}}. \bibinfo{pages}{515--524}.
\newblock


\bibitem[Jordan(2003)]%
        {jordan2003introduction}
\bibfield{author}{\bibinfo{person}{Michael~I Jordan}.} \bibinfo{year}{2003}\natexlab{}.
\newblock \bibinfo{title}{An introduction to probabilistic graphical models}.
\newblock
\newblock


\bibitem[Josiam and Hobson(1995)]%
        {josiam1995consumer}
\bibfield{author}{\bibinfo{person}{Bharath~M Josiam} {and} \bibinfo{person}{JS~Perry Hobson}.} \bibinfo{year}{1995}\natexlab{}.
\newblock \showarticletitle{Consumer choice in context: the decoy effect in travel and tourism}.
\newblock \bibinfo{journal}{\emph{Journal of Travel Research}} \bibinfo{volume}{34}, \bibinfo{number}{1} (\bibinfo{year}{1995}), \bibinfo{pages}{45--50}.
\newblock


\bibitem[Kang and McAuley(2018)]%
        {kang2018self}
\bibfield{author}{\bibinfo{person}{Wang-Cheng Kang} {and} \bibinfo{person}{Julian McAuley}.} \bibinfo{year}{2018}\natexlab{}.
\newblock \showarticletitle{Self-attentive sequential recommendation}. In \bibinfo{booktitle}{\emph{2018 IEEE International Conference on Data Mining (ICDM)}}. IEEE, \bibinfo{pages}{197--206}.
\newblock


\bibitem[Koren(2008)]%
        {koren2008factorization}
\bibfield{author}{\bibinfo{person}{Yehuda Koren}.} \bibinfo{year}{2008}\natexlab{}.
\newblock \showarticletitle{Factorization meets the neighborhood: a multifaceted collaborative filtering model}. In \bibinfo{booktitle}{\emph{Proceedings of the 14th ACM SIGKDD international conference on Knowledge discovery and data mining}}. \bibinfo{pages}{426--434}.
\newblock


\bibitem[Krumm et~al\mbox{.}(2008)]%
        {krumm2008user}
\bibfield{author}{\bibinfo{person}{John Krumm}, \bibinfo{person}{Nigel Davies}, {and} \bibinfo{person}{Chandra Narayanaswami}.} \bibinfo{year}{2008}\natexlab{}.
\newblock \showarticletitle{User-generated content}.
\newblock \bibinfo{journal}{\emph{IEEE Pervasive Computing}} \bibinfo{volume}{7}, \bibinfo{number}{4} (\bibinfo{year}{2008}), \bibinfo{pages}{10--11}.
\newblock


\bibitem[Lemon et~al\mbox{.}(2006)]%
        {lemon2006evaluating}
\bibfield{author}{\bibinfo{person}{Oliver Lemon}, \bibinfo{person}{Kallirroi Georgila}, {and} \bibinfo{person}{James Henderson}.} \bibinfo{year}{2006}\natexlab{}.
\newblock \showarticletitle{Evaluating effectiveness and portability of reinforcement learned dialogue strategies with real users: the TALK TownInfo evaluation}. In \bibinfo{booktitle}{\emph{2006 IEEE Spoken Language Technology Workshop}}. IEEE, \bibinfo{pages}{178--181}.
\newblock


\bibitem[Levin et~al\mbox{.}(2000)]%
        {levin2000stochastic}
\bibfield{author}{\bibinfo{person}{Esther Levin}, \bibinfo{person}{Roberto Pieraccini}, {and} \bibinfo{person}{Wieland Eckert}.} \bibinfo{year}{2000}\natexlab{}.
\newblock \showarticletitle{A stochastic model of human-machine interaction for learning dialog strategies}.
\newblock \bibinfo{journal}{\emph{IEEE Transactions on speech and audio processing}} \bibinfo{volume}{8}, \bibinfo{number}{1} (\bibinfo{year}{2000}), \bibinfo{pages}{11--23}.
\newblock


\bibitem[Lex et~al\mbox{.}(2018)]%
        {lex2018mitigating}
\bibfield{author}{\bibinfo{person}{Elisabeth Lex}, \bibinfo{person}{Mario Wagner}, {and} \bibinfo{person}{Dominik Kowald}.} \bibinfo{year}{2018}\natexlab{}.
\newblock \showarticletitle{Mitigating confirmation bias on twitter by recommending opposing views}.
\newblock \bibinfo{journal}{\emph{arXiv preprint arXiv:1809.03901}} (\bibinfo{year}{2018}).
\newblock


\bibitem[Li et~al\mbox{.}(2014)]%
        {li2014modeling}
\bibfield{author}{\bibinfo{person}{Lei Li}, \bibinfo{person}{Li Zheng}, \bibinfo{person}{Fan Yang}, {and} \bibinfo{person}{Tao Li}.} \bibinfo{year}{2014}\natexlab{}.
\newblock \showarticletitle{Modeling and broadening temporal user interest in personalized news recommendation}.
\newblock \bibinfo{journal}{\emph{Expert Systems with Applications}} \bibinfo{volume}{41}, \bibinfo{number}{7} (\bibinfo{year}{2014}), \bibinfo{pages}{3168--3177}.
\newblock


\bibitem[Liu et~al\mbox{.}(2019)]%
        {liu2019user}
\bibfield{author}{\bibinfo{person}{Shang Liu}, \bibinfo{person}{Zhenzhong Chen}, \bibinfo{person}{Hongyi Liu}, {and} \bibinfo{person}{Xinghai Hu}.} \bibinfo{year}{2019}\natexlab{}.
\newblock \showarticletitle{User-video co-attention network for personalized micro-video recommendation}. In \bibinfo{booktitle}{\emph{The World Wide Web Conference}}. \bibinfo{pages}{3020--3026}.
\newblock


\bibitem[Mansoury et~al\mbox{.}(2020)]%
        {mansoury2020feedback}
\bibfield{author}{\bibinfo{person}{Masoud Mansoury}, \bibinfo{person}{Himan Abdollahpouri}, \bibinfo{person}{Mykola Pechenizkiy}, \bibinfo{person}{Bamshad Mobasher}, {and} \bibinfo{person}{Robin Burke}.} \bibinfo{year}{2020}\natexlab{}.
\newblock \showarticletitle{Feedback loop and bias amplification in recommender systems}. In \bibinfo{booktitle}{\emph{Proceedings of the 29th ACM international conference on information \& knowledge management}}. \bibinfo{pages}{2145--2148}.
\newblock


\bibitem[Milano et~al\mbox{.}(2020)]%
        {milano2020recommender}
\bibfield{author}{\bibinfo{person}{Silvia Milano}, \bibinfo{person}{Mariarosaria Taddeo}, {and} \bibinfo{person}{Luciano Floridi}.} \bibinfo{year}{2020}\natexlab{}.
\newblock \showarticletitle{Recommender systems and their ethical challenges}.
\newblock \bibinfo{journal}{\emph{Ai \& Society}} \bibinfo{volume}{35}, \bibinfo{number}{4} (\bibinfo{year}{2020}), \bibinfo{pages}{957--967}.
\newblock


\bibitem[Min(2003)]%
        {min2003consumer}
\bibfield{author}{\bibinfo{person}{Kyeong~Sam Min}.} \bibinfo{year}{2003}\natexlab{}.
\newblock \bibinfo{booktitle}{\emph{Consumer response to product unavailability}}.
\newblock \bibinfo{publisher}{The Ohio State University}.
\newblock


\bibitem[Mladenov et~al\mbox{.}(2020)]%
        {mladenov2020demonstrating}
\bibfield{author}{\bibinfo{person}{Martin Mladenov}, \bibinfo{person}{Chih-wei Hsu}, \bibinfo{person}{Vihan Jain}, \bibinfo{person}{Eugene Ie}, \bibinfo{person}{Christopher Colby}, \bibinfo{person}{Nicolas Mayoraz}, \bibinfo{person}{Hubert Pham}, \bibinfo{person}{Dustin Tran}, \bibinfo{person}{Ivan Vendrov}, {and} \bibinfo{person}{Craig Boutilier}.} \bibinfo{year}{2020}\natexlab{}.
\newblock \showarticletitle{Demonstrating Principled Uncertainty Modeling for Recommender Ecosystems with RecSim NG}. In \bibinfo{booktitle}{\emph{Fourteenth ACM Conference on Recommender Systems}}. \bibinfo{pages}{591--593}.
\newblock


\bibitem[Mnih et~al\mbox{.}(2013)]%
        {mnih2013playing}
\bibfield{author}{\bibinfo{person}{Volodymyr Mnih}, \bibinfo{person}{Koray Kavukcuoglu}, \bibinfo{person}{David Silver}, \bibinfo{person}{Alex Graves}, \bibinfo{person}{Ioannis Antonoglou}, \bibinfo{person}{Daan Wierstra}, {and} \bibinfo{person}{Martin Riedmiller}.} \bibinfo{year}{2013}\natexlab{}.
\newblock \showarticletitle{Playing atari with deep reinforcement learning}.
\newblock \bibinfo{journal}{\emph{arXiv preprint arXiv:1312.5602}} (\bibinfo{year}{2013}).
\newblock


\bibitem[Nguyen et~al\mbox{.}(2014)]%
        {nguyen2014exploring}
\bibfield{author}{\bibinfo{person}{Tien~T Nguyen}, \bibinfo{person}{Pik-Mai Hui}, \bibinfo{person}{F~Maxwell Harper}, \bibinfo{person}{Loren Terveen}, {and} \bibinfo{person}{Joseph~A Konstan}.} \bibinfo{year}{2014}\natexlab{}.
\newblock \showarticletitle{Exploring the filter bubble: the effect of using recommender systems on content diversity}. In \bibinfo{booktitle}{\emph{Proceedings of the 23rd international conference on World wide web}}. \bibinfo{pages}{677--686}.
\newblock


\bibitem[Nickerson(1998)]%
        {nickerson1998confirmation}
\bibfield{author}{\bibinfo{person}{Raymond~S Nickerson}.} \bibinfo{year}{1998}\natexlab{}.
\newblock \showarticletitle{Confirmation bias: A ubiquitous phenomenon in many guises}.
\newblock \bibinfo{journal}{\emph{Review of general psychology}} \bibinfo{volume}{2}, \bibinfo{number}{2} (\bibinfo{year}{1998}), \bibinfo{pages}{175--220}.
\newblock


\bibitem[Pan et~al\mbox{.}(2022)]%
        {pan2022exploiting}
\bibfield{author}{\bibinfo{person}{Xiao Pan}, \bibinfo{person}{Lei Wu}, \bibinfo{person}{Fenjie Long}, {and} \bibinfo{person}{Ang Ma}.} \bibinfo{year}{2022}\natexlab{}.
\newblock \showarticletitle{Exploiting user behavior learning for personalized trajectory recommendations}.
\newblock \bibinfo{journal}{\emph{Frontiers of Computer Science}}  \bibinfo{volume}{16} (\bibinfo{year}{2022}), \bibinfo{pages}{1--12}.
\newblock


\bibitem[Pang et~al\mbox{.}(2020)]%
        {pang2020setrank}
\bibfield{author}{\bibinfo{person}{Liang Pang}, \bibinfo{person}{Jun Xu}, \bibinfo{person}{Qingyao Ai}, \bibinfo{person}{Yanyan Lan}, \bibinfo{person}{Xueqi Cheng}, {and} \bibinfo{person}{Jirong Wen}.} \bibinfo{year}{2020}\natexlab{}.
\newblock \showarticletitle{Setrank: Learning a permutation-invariant ranking model for information retrieval}. In \bibinfo{booktitle}{\emph{Proceedings of the 43rd International ACM SIGIR Conference on Research and Development in Information Retrieval}}. \bibinfo{pages}{499--508}.
\newblock


\bibitem[Paraschakis(2017)]%
        {paraschakis2017towards}
\bibfield{author}{\bibinfo{person}{Dimitris Paraschakis}.} \bibinfo{year}{2017}\natexlab{}.
\newblock \showarticletitle{Towards an ethical recommendation framework}. In \bibinfo{booktitle}{\emph{2017 11th international conference on research challenges in information science (RCIS)}}. IEEE, \bibinfo{pages}{211--220}.
\newblock


\bibitem[Pei et~al\mbox{.}(2019)]%
        {pei2019personalized}
\bibfield{author}{\bibinfo{person}{Changhua Pei}, \bibinfo{person}{Yi Zhang}, \bibinfo{person}{Yongfeng Zhang}, \bibinfo{person}{Fei Sun}, \bibinfo{person}{Xiao Lin}, \bibinfo{person}{Hanxiao Sun}, \bibinfo{person}{Jian Wu}, \bibinfo{person}{Peng Jiang}, \bibinfo{person}{Junfeng Ge}, \bibinfo{person}{Wenwu Ou}, {et~al\mbox{.}}} \bibinfo{year}{2019}\natexlab{}.
\newblock \showarticletitle{Personalized re-ranking for recommendation}. In \bibinfo{booktitle}{\emph{Proceedings of the 13th ACM conference on recommender systems}}. \bibinfo{pages}{3--11}.
\newblock


\bibitem[Pietquin and Dutoit(2006)]%
        {pietquin2006probabilistic}
\bibfield{author}{\bibinfo{person}{Olivier Pietquin} {and} \bibinfo{person}{Thierry Dutoit}.} \bibinfo{year}{2006}\natexlab{}.
\newblock \showarticletitle{A probabilistic framework for dialog simulation and optimal strategy learning}.
\newblock \bibinfo{journal}{\emph{IEEE Transactions on Audio, Speech, and Language Processing}} \bibinfo{volume}{14}, \bibinfo{number}{2} (\bibinfo{year}{2006}), \bibinfo{pages}{589--599}.
\newblock


\bibitem[Qin et~al\mbox{.}(2020)]%
        {qin2020user}
\bibfield{author}{\bibinfo{person}{Jiarui Qin}, \bibinfo{person}{Weinan Zhang}, \bibinfo{person}{Xin Wu}, \bibinfo{person}{Jiarui Jin}, \bibinfo{person}{Yuchen Fang}, {and} \bibinfo{person}{Yong Yu}.} \bibinfo{year}{2020}\natexlab{}.
\newblock \showarticletitle{User behavior retrieval for click-through rate prediction}. In \bibinfo{booktitle}{\emph{Proceedings of the 43rd International ACM SIGIR Conference on Research and Development in Information Retrieval}}. \bibinfo{pages}{2347--2356}.
\newblock


\bibitem[Ren et~al\mbox{.}(2019)]%
        {ren2019lifelong}
\bibfield{author}{\bibinfo{person}{Kan Ren}, \bibinfo{person}{Jiarui Qin}, \bibinfo{person}{Yuchen Fang}, \bibinfo{person}{Weinan Zhang}, \bibinfo{person}{Lei Zheng}, \bibinfo{person}{Weijie Bian}, \bibinfo{person}{Guorui Zhou}, \bibinfo{person}{Jian Xu}, \bibinfo{person}{Yong Yu}, \bibinfo{person}{Xiaoqiang Zhu}, {et~al\mbox{.}}} \bibinfo{year}{2019}\natexlab{}.
\newblock \showarticletitle{Lifelong sequential modeling with personalized memorization for user response prediction}. In \bibinfo{booktitle}{\emph{Proceedings of the 42nd International ACM SIGIR Conference on Research and Development in Information Retrieval}}. \bibinfo{pages}{565--574}.
\newblock


\bibitem[Rendle et~al\mbox{.}(2010)]%
        {rendle2010factorizing}
\bibfield{author}{\bibinfo{person}{Steffen Rendle}, \bibinfo{person}{Christoph Freudenthaler}, {and} \bibinfo{person}{Lars Schmidt-Thieme}.} \bibinfo{year}{2010}\natexlab{}.
\newblock \showarticletitle{Factorizing personalized markov chains for next-basket recommendation}. In \bibinfo{booktitle}{\emph{Proceedings of the 19th international conference on World wide web}}. \bibinfo{pages}{811--820}.
\newblock


\bibitem[Ribeiro et~al\mbox{.}(2020)]%
        {ribeiro2020auditing}
\bibfield{author}{\bibinfo{person}{Manoel~Horta Ribeiro}, \bibinfo{person}{Raphael Ottoni}, \bibinfo{person}{Robert West}, \bibinfo{person}{Virg{\'\i}lio~AF Almeida}, {and} \bibinfo{person}{Wagner Meira~Jr}.} \bibinfo{year}{2020}\natexlab{}.
\newblock \showarticletitle{Auditing radicalization pathways on YouTube}. In \bibinfo{booktitle}{\emph{Proceedings of the 2020 conference on fairness, accountability, and transparency}}. \bibinfo{pages}{131--141}.
\newblock


\bibitem[Rohde et~al\mbox{.}(2018)]%
        {rohde2018recogym}
\bibfield{author}{\bibinfo{person}{David Rohde}, \bibinfo{person}{Stephen Bonner}, \bibinfo{person}{Travis Dunlop}, \bibinfo{person}{Flavian Vasile}, {and} \bibinfo{person}{Alexandros Karatzoglou}.} \bibinfo{year}{2018}\natexlab{}.
\newblock \showarticletitle{Recogym: A reinforcement learning environment for the problem of product recommendation in online advertising}.
\newblock \bibinfo{journal}{\emph{arXiv preprint arXiv:1808.00720}} (\bibinfo{year}{2018}).
\newblock


\bibitem[Sarwar et~al\mbox{.}(2001)]%
        {sarwar2001item}
\bibfield{author}{\bibinfo{person}{Badrul Sarwar}, \bibinfo{person}{George Karypis}, \bibinfo{person}{Joseph Konstan}, {and} \bibinfo{person}{John Riedl}.} \bibinfo{year}{2001}\natexlab{}.
\newblock \showarticletitle{Item-based collaborative filtering recommendation algorithms}. In \bibinfo{booktitle}{\emph{Proceedings of the 10th international conference on World Wide Web}}. \bibinfo{pages}{285--295}.
\newblock


\bibitem[Schatzmann et~al\mbox{.}(2006)]%
        {schatzmann2006survey}
\bibfield{author}{\bibinfo{person}{Jost Schatzmann}, \bibinfo{person}{Karl Weilhammer}, \bibinfo{person}{Matt Stuttle}, {and} \bibinfo{person}{Steve Young}.} \bibinfo{year}{2006}\natexlab{}.
\newblock \showarticletitle{A survey of statistical user simulation techniques for reinforcement-learning of dialogue management strategies}.
\newblock \bibinfo{journal}{\emph{The knowledge engineering review}} \bibinfo{volume}{21}, \bibinfo{number}{2} (\bibinfo{year}{2006}), \bibinfo{pages}{97--126}.
\newblock


\bibitem[Schoemaker(1982)]%
        {schoemaker1982expected}
\bibfield{author}{\bibinfo{person}{Paul~JH Schoemaker}.} \bibinfo{year}{1982}\natexlab{}.
\newblock \showarticletitle{The expected utility model: Its variants, purposes, evidence and limitations}.
\newblock \bibinfo{journal}{\emph{Journal of economic literature}} (\bibinfo{year}{1982}), \bibinfo{pages}{529--563}.
\newblock


\bibitem[Schulman et~al\mbox{.}(2017)]%
        {schulman2017proximal}
\bibfield{author}{\bibinfo{person}{John Schulman}, \bibinfo{person}{Filip Wolski}, \bibinfo{person}{Prafulla Dhariwal}, \bibinfo{person}{Alec Radford}, {and} \bibinfo{person}{Oleg Klimov}.} \bibinfo{year}{2017}\natexlab{}.
\newblock \showarticletitle{Proximal policy optimization algorithms}.
\newblock \bibinfo{journal}{\emph{arXiv preprint arXiv:1707.06347}} (\bibinfo{year}{2017}).
\newblock


\bibitem[Selten(1990)]%
        {selten1990bounded}
\bibfield{author}{\bibinfo{person}{Reinhard Selten}.} \bibinfo{year}{1990}\natexlab{}.
\newblock \showarticletitle{Bounded rationality}.
\newblock \bibinfo{journal}{\emph{Journal of Institutional and Theoretical Economics (JITE)/Zeitschrift f{\"u}r die gesamte Staatswissenschaft}} \bibinfo{volume}{146}, \bibinfo{number}{4} (\bibinfo{year}{1990}), \bibinfo{pages}{649--658}.
\newblock


\bibitem[Shang et~al\mbox{.}(2019)]%
        {shang2019environment}
\bibfield{author}{\bibinfo{person}{Wenjie Shang}, \bibinfo{person}{Yang Yu}, \bibinfo{person}{Qingyang Li}, \bibinfo{person}{Zhiwei Qin}, \bibinfo{person}{Yiping Meng}, {and} \bibinfo{person}{Jieping Ye}.} \bibinfo{year}{2019}\natexlab{}.
\newblock \showarticletitle{Environment reconstruction with hidden confounders for reinforcement learning based recommendation}. In \bibinfo{booktitle}{\emph{Proceedings of the 25th ACM SIGKDD International Conference on Knowledge Discovery \& Data Mining}}. \bibinfo{pages}{566--576}.
\newblock


\bibitem[Shi et~al\mbox{.}(2019a)]%
        {shi2019pyrecgym}
\bibfield{author}{\bibinfo{person}{Bichen Shi}, \bibinfo{person}{Makbule~Gulcin Ozsoy}, \bibinfo{person}{Neil Hurley}, \bibinfo{person}{Barry Smyth}, \bibinfo{person}{Elias~Z Tragos}, \bibinfo{person}{James Geraci}, {and} \bibinfo{person}{Aonghus Lawlor}.} \bibinfo{year}{2019}\natexlab{a}.
\newblock \showarticletitle{Pyrecgym: A reinforcement learning gym for recommender systems}. In \bibinfo{booktitle}{\emph{Proceedings of the 13th ACM Conference on Recommender Systems}}. \bibinfo{pages}{491--495}.
\newblock


\bibitem[Shi et~al\mbox{.}(2019b)]%
        {shi2019virtual}
\bibfield{author}{\bibinfo{person}{Jing-Cheng Shi}, \bibinfo{person}{Yang Yu}, \bibinfo{person}{Qing Da}, \bibinfo{person}{Shi-Yong Chen}, {and} \bibinfo{person}{An-Xiang Zeng}.} \bibinfo{year}{2019}\natexlab{b}.
\newblock \showarticletitle{Virtual-taobao: Virtualizing real-world online retail environment for reinforcement learning}. In \bibinfo{booktitle}{\emph{Proceedings of the AAAI Conference on Artificial Intelligence}}, Vol.~\bibinfo{volume}{33}. \bibinfo{pages}{4902--4909}.
\newblock


\bibitem[Slaughter et~al\mbox{.}(1999)]%
        {slaughter1999decoy}
\bibfield{author}{\bibinfo{person}{Jerel~E Slaughter}, \bibinfo{person}{Evan~F Sinar}, {and} \bibinfo{person}{Scott Highhouse}.} \bibinfo{year}{1999}\natexlab{}.
\newblock \showarticletitle{Decoy effects and attribute-level inferences.}
\newblock \bibinfo{journal}{\emph{Journal of applied psychology}} \bibinfo{volume}{84}, \bibinfo{number}{5} (\bibinfo{year}{1999}), \bibinfo{pages}{823}.
\newblock


\bibitem[Srikant et~al\mbox{.}(2010)]%
        {srikant2010user}
\bibfield{author}{\bibinfo{person}{Ramakrishnan Srikant}, \bibinfo{person}{Sugato Basu}, \bibinfo{person}{Ni Wang}, {and} \bibinfo{person}{Daryl Pregibon}.} \bibinfo{year}{2010}\natexlab{}.
\newblock \showarticletitle{User browsing models: relevance versus examination}. In \bibinfo{booktitle}{\emph{Proceedings of the 16th ACM SIGKDD international conference on Knowledge discovery and data mining}}. \bibinfo{pages}{223--232}.
\newblock


\bibitem[Sun et~al\mbox{.}(2019)]%
        {sun2019bert4rec}
\bibfield{author}{\bibinfo{person}{Fei Sun}, \bibinfo{person}{Jun Liu}, \bibinfo{person}{Jian Wu}, \bibinfo{person}{Changhua Pei}, \bibinfo{person}{Xiao Lin}, \bibinfo{person}{Wenwu Ou}, {and} \bibinfo{person}{Peng Jiang}.} \bibinfo{year}{2019}\natexlab{}.
\newblock \showarticletitle{BERT4Rec: Sequential recommendation with bidirectional encoder representations from transformer}. In \bibinfo{booktitle}{\emph{Proceedings of the 28th ACM international conference on information and knowledge management}}. \bibinfo{pages}{1441--1450}.
\newblock


\bibitem[Tang and Wang(2018)]%
        {tang2018personalized}
\bibfield{author}{\bibinfo{person}{Jiaxi Tang} {and} \bibinfo{person}{Ke Wang}.} \bibinfo{year}{2018}\natexlab{}.
\newblock \showarticletitle{Personalized top-n sequential recommendation via convolutional sequence embedding}. In \bibinfo{booktitle}{\emph{Proceedings of the eleventh ACM international conference on web search and data mining}}. \bibinfo{pages}{565--573}.
\newblock


\bibitem[Teppan et~al\mbox{.}(2011)]%
        {teppan2011decoy}
\bibfield{author}{\bibinfo{person}{Erich Teppan}, \bibinfo{person}{Alexander Felfernig}, {and} \bibinfo{person}{Klaus Isak}.} \bibinfo{year}{2011}\natexlab{}.
\newblock \showarticletitle{Decoy effects in financial service e-sales systems}. In \bibinfo{booktitle}{\emph{Proceedings of the Workshop Decisions@ RecSys, in Conjunction with the Fourth ACM Conference on Recommender Systems}}. Citeseer, \bibinfo{pages}{1--8}.
\newblock


\bibitem[Tian and Ekstrand(2020)]%
        {tian2020estimating}
\bibfield{author}{\bibinfo{person}{Mucun Tian} {and} \bibinfo{person}{Michael~D Ekstrand}.} \bibinfo{year}{2020}\natexlab{}.
\newblock \showarticletitle{Estimating error and bias in offline evaluation results}. In \bibinfo{booktitle}{\emph{Proceedings of the 2020 Conference on Human Information Interaction and Retrieval}}. \bibinfo{pages}{392--396}.
\newblock


\bibitem[Wang et~al\mbox{.}(2018b)]%
        {wang2018dkn}
\bibfield{author}{\bibinfo{person}{Hongwei Wang}, \bibinfo{person}{Fuzheng Zhang}, \bibinfo{person}{Xing Xie}, {and} \bibinfo{person}{Minyi Guo}.} \bibinfo{year}{2018}\natexlab{b}.
\newblock \showarticletitle{DKN: Deep knowledge-aware network for news recommendation}. In \bibinfo{booktitle}{\emph{Proceedings of the 2018 world wide web conference}}. \bibinfo{pages}{1835--1844}.
\newblock


\bibitem[Wang et~al\mbox{.}(2021)]%
        {wang2021rl4rs}
\bibfield{author}{\bibinfo{person}{Kai Wang}, \bibinfo{person}{Zhene Zou}, \bibinfo{person}{Qilin Deng}, \bibinfo{person}{Yue Shang}, \bibinfo{person}{Minghao Zhao}, \bibinfo{person}{Runze Wu}, \bibinfo{person}{Xudong Shen}, \bibinfo{person}{Tangjie Lyu}, {and} \bibinfo{person}{Changjie Fan}.} \bibinfo{year}{2021}\natexlab{}.
\newblock \showarticletitle{Rl4rs: A real-world benchmark for reinforcement learning based recommender system}.
\newblock \bibinfo{journal}{\emph{arXiv preprint arXiv:2110.11073}} (\bibinfo{year}{2021}).
\newblock


\bibitem[Wang and Chen(2021)]%
        {wang2021user}
\bibfield{author}{\bibinfo{person}{Ningxia Wang} {and} \bibinfo{person}{Li Chen}.} \bibinfo{year}{2021}\natexlab{}.
\newblock \showarticletitle{User Bias in Beyond-Accuracy Measurement of Recommendation Algorithms}. In \bibinfo{booktitle}{\emph{Fifteenth ACM Conference on Recommender Systems}}. \bibinfo{pages}{133--142}.
\newblock


\bibitem[Wang et~al\mbox{.}(2018a)]%
        {wang2018lambdaloss}
\bibfield{author}{\bibinfo{person}{Xuanhui Wang}, \bibinfo{person}{Cheng Li}, \bibinfo{person}{Nadav Golbandi}, \bibinfo{person}{Michael Bendersky}, {and} \bibinfo{person}{Marc Najork}.} \bibinfo{year}{2018}\natexlab{a}.
\newblock \showarticletitle{The lambdaloss framework for ranking metric optimization}. In \bibinfo{booktitle}{\emph{Proceedings of the 27th ACM international conference on information and knowledge management}}. \bibinfo{pages}{1313--1322}.
\newblock


\bibitem[Wang et~al\mbox{.}(2022)]%
        {wang2022survey}
\bibfield{author}{\bibinfo{person}{Yifan Wang}, \bibinfo{person}{Weizhi Ma}, \bibinfo{person}{Min Zhang*}, \bibinfo{person}{Yiqun Liu}, {and} \bibinfo{person}{Shaoping Ma}.} \bibinfo{year}{2022}\natexlab{}.
\newblock \showarticletitle{A Survey on the Fairness of Recommender Systems}.
\newblock \bibinfo{journal}{\emph{ACM Journal of the ACM (JACM)}} (\bibinfo{year}{2022}).
\newblock


\bibitem[Webber et~al\mbox{.}(2010)]%
        {webber2010similarity}
\bibfield{author}{\bibinfo{person}{William Webber}, \bibinfo{person}{Alistair Moffat}, {and} \bibinfo{person}{Justin Zobel}.} \bibinfo{year}{2010}\natexlab{}.
\newblock \showarticletitle{A similarity measure for indefinite rankings}.
\newblock \bibinfo{journal}{\emph{ACM Transactions on Information Systems (TOIS)}} \bibinfo{volume}{28}, \bibinfo{number}{4} (\bibinfo{year}{2010}), \bibinfo{pages}{1--38}.
\newblock


\bibitem[Xia et~al\mbox{.}(2008)]%
        {xia2008listwise}
\bibfield{author}{\bibinfo{person}{Fen Xia}, \bibinfo{person}{Tie-Yan Liu}, \bibinfo{person}{Jue Wang}, \bibinfo{person}{Wensheng Zhang}, {and} \bibinfo{person}{Hang Li}.} \bibinfo{year}{2008}\natexlab{}.
\newblock \showarticletitle{Listwise approach to learning to rank: theory and algorithm}. In \bibinfo{booktitle}{\emph{Proceedings of the 25th international conference on Machine learning}}. \bibinfo{pages}{1192--1199}.
\newblock


\bibitem[Xiao and Wang(2021)]%
        {xiao2021general}
\bibfield{author}{\bibinfo{person}{Teng Xiao} {and} \bibinfo{person}{Donglin Wang}.} \bibinfo{year}{2021}\natexlab{}.
\newblock \showarticletitle{A general offline reinforcement learning framework for interactive recommendation}. In \bibinfo{booktitle}{\emph{The Thirty-Fifth AAAI Conference on Artificial Intelligence, AAAI}}, Vol.~\bibinfo{volume}{2021}.
\newblock


\bibitem[Yao et~al\mbox{.}(2021)]%
        {yao2021measuring}
\bibfield{author}{\bibinfo{person}{Sirui Yao}, \bibinfo{person}{Yoni Halpern}, \bibinfo{person}{Nithum Thain}, \bibinfo{person}{Xuezhi Wang}, \bibinfo{person}{Kang Lee}, \bibinfo{person}{Flavien Prost}, \bibinfo{person}{Ed~H Chi}, \bibinfo{person}{Jilin Chen}, {and} \bibinfo{person}{Alex Beutel}.} \bibinfo{year}{2021}\natexlab{}.
\newblock \showarticletitle{Measuring recommender system effects with simulated users}.
\newblock \bibinfo{journal}{\emph{arXiv preprint arXiv:2101.04526}} (\bibinfo{year}{2021}).
\newblock


\bibitem[Yuan et~al\mbox{.}(2016)]%
        {yuan2016lambdafm}
\bibfield{author}{\bibinfo{person}{Fajie Yuan}, \bibinfo{person}{Guibing Guo}, \bibinfo{person}{Joemon~M Jose}, \bibinfo{person}{Long Chen}, \bibinfo{person}{Haitao Yu}, {and} \bibinfo{person}{Weinan Zhang}.} \bibinfo{year}{2016}\natexlab{}.
\newblock \showarticletitle{Lambdafm: learning optimal ranking with factorization machines using lambda surrogates}. In \bibinfo{booktitle}{\emph{Proceedings of the 25th ACM international on conference on information and knowledge management}}. \bibinfo{pages}{227--236}.
\newblock


\bibitem[Zhang et~al\mbox{.}(2021)]%
        {zhang2021survey}
\bibfield{author}{\bibinfo{person}{Yu Zhang}, \bibinfo{person}{Peter Ti{\v{n}}o}, \bibinfo{person}{Ale{\v{s}} Leonardis}, {and} \bibinfo{person}{Ke Tang}.} \bibinfo{year}{2021}\natexlab{}.
\newblock \showarticletitle{A survey on neural network interpretability}.
\newblock \bibinfo{journal}{\emph{IEEE Transactions on Emerging Topics in Computational Intelligence}} (\bibinfo{year}{2021}).
\newblock


\bibitem[Zhao et~al\mbox{.}(2019)]%
        {zhao2019recommending}
\bibfield{author}{\bibinfo{person}{Zhe Zhao}, \bibinfo{person}{Lichan Hong}, \bibinfo{person}{Li Wei}, \bibinfo{person}{Jilin Chen}, \bibinfo{person}{Aniruddh Nath}, \bibinfo{person}{Shawn Andrews}, \bibinfo{person}{Aditee Kumthekar}, \bibinfo{person}{Maheswaran Sathiamoorthy}, \bibinfo{person}{Xinyang Yi}, {and} \bibinfo{person}{Ed Chi}.} \bibinfo{year}{2019}\natexlab{}.
\newblock \showarticletitle{Recommending what video to watch next: a multitask ranking system}. In \bibinfo{booktitle}{\emph{Proceedings of the 13th ACM Conference on Recommender Systems}}. \bibinfo{pages}{43--51}.
\newblock


\bibitem[Zhou et~al\mbox{.}(2019)]%
        {zhou2019deep}
\bibfield{author}{\bibinfo{person}{Guorui Zhou}, \bibinfo{person}{Na Mou}, \bibinfo{person}{Ying Fan}, \bibinfo{person}{Qi Pi}, \bibinfo{person}{Weijie Bian}, \bibinfo{person}{Chang Zhou}, \bibinfo{person}{Xiaoqiang Zhu}, {and} \bibinfo{person}{Kun Gai}.} \bibinfo{year}{2019}\natexlab{}.
\newblock \showarticletitle{Deep interest evolution network for click-through rate prediction}. In \bibinfo{booktitle}{\emph{Proceedings of the AAAI conference on artificial intelligence}}, Vol.~\bibinfo{volume}{33}. \bibinfo{pages}{5941--5948}.
\newblock


\bibitem[Zhou et~al\mbox{.}(2018)]%
        {zhou2018deep}
\bibfield{author}{\bibinfo{person}{Guorui Zhou}, \bibinfo{person}{Xiaoqiang Zhu}, \bibinfo{person}{Chenru Song}, \bibinfo{person}{Ying Fan}, \bibinfo{person}{Han Zhu}, \bibinfo{person}{Xiao Ma}, \bibinfo{person}{Yanghui Yan}, \bibinfo{person}{Junqi Jin}, \bibinfo{person}{Han Li}, {and} \bibinfo{person}{Kun Gai}.} \bibinfo{year}{2018}\natexlab{}.
\newblock \showarticletitle{Deep interest network for click-through rate prediction}. In \bibinfo{booktitle}{\emph{Proceedings of the 24th ACM SIGKDD international conference on knowledge discovery \& data mining}}. \bibinfo{pages}{1059--1068}.
\newblock


\bibitem[Zhu et~al\mbox{.}(2021)]%
        {zhu2021popularity}
\bibfield{author}{\bibinfo{person}{Ziwei Zhu}, \bibinfo{person}{Yun He}, \bibinfo{person}{Xing Zhao}, \bibinfo{person}{Yin Zhang}, \bibinfo{person}{Jianling Wang}, {and} \bibinfo{person}{James Caverlee}.} \bibinfo{year}{2021}\natexlab{}.
\newblock \showarticletitle{Popularity-opportunity bias in collaborative filtering}. In \bibinfo{booktitle}{\emph{Proceedings of the 14th ACM International Conference on Web Search and Data Mining}}. \bibinfo{pages}{85--93}.
\newblock


\bibitem[Zhu et~al\mbox{.}(2020)]%
        {zhu2020measuring}
\bibfield{author}{\bibinfo{person}{Ziwei Zhu}, \bibinfo{person}{Jianling Wang}, {and} \bibinfo{person}{James Caverlee}.} \bibinfo{year}{2020}\natexlab{}.
\newblock \showarticletitle{Measuring and mitigating item under-recommendation bias in personalized ranking systems}. In \bibinfo{booktitle}{\emph{Proceedings of the 43rd international ACM SIGIR conference on research and development in information retrieval}}. \bibinfo{pages}{449--458}.
\newblock


\end{thebibliography}

\end{document}